\documentclass[runningheads,a4paper]{llncs}

\usepackage[leqno,tbtags,fleqn]{amsmath}
\usepackage{amssymb}
\usepackage{theorem}
\usepackage{slashed}
\usepackage{fancyvrb}

\usepackage{graphicx}
\usepackage{url}
\usepackage{moreverb} 
\usepackage{wrapfig}

\usepackage[latin1]{inputenc}
\usepackage[active]{srcltx} 

\usepackage{graphicx}
\usepackage{xcolor}
\usepackage{wrapfig}
\usepackage{titletoc}
\usepackage{tocloft}

\newcommand{\hpilot}{\mbox{H-PILoT} }

\newcommand{\T}{{\mathcal T}}

\newcommand{\falsum}{\bot}

\newcommand{\x}{\bar{x}}

\newcommand{\K}{{\mathcal K}}

\newenvironment{enumerate-} 
{\begin{enumerate}
   \setlength{\parskip}{-1ex}              
   \setlength{\itemsep}{1.5ex}             
}
{
 \end{enumerate}
}

\newenvironment{itemize-} 
{\begin{itemize}
   \setlength{\parskip}{-1ex}              
   \setlength{\itemsep}{1.5ex}             
}
{
 \end{itemize}
}

\DefineVerbatimEnvironment%
{MyVerbatim}{Verbatim}
{fontsize=\small}

\DefineVerbatimEnvironment%
{MySmallVerbatim}{Verbatim}
{fontsize=\tiny}

\DefineVerbatimEnvironment%
{MyFVerbatim}{Verbatim}
{fontsize=\scriptsize,frame=single,framesep=2mm}

\begin{document}

\title{System Description: \hpilot \\ Version 1.9}
\titlerunning{System Description: H-PILoT}

\author{Carsten Ihlemann and Viorica Sofronie-Stokkermans}
\authorrunning{Carsten Ihlemann and Viorica Sofronie-Stokkermans}

\institute{Max-Planck-Institut f{\"u}r Informatik\\
Campus E1.4, Saarbr{\"u}cken\\
e-mail: {\tt \{ihlemann|sofronie\}@mpi-inf.mpg.de} }

\maketitle

\begin{abstract}
This system description provides an overview of H-PILoT 
(Hierarchical Proving by Instantiation in Local Theory extensions),  
a program for hierarchical reasoning in extensions of logical theories. 
H-PILoT reduces deduction problems in the theory extension to deduction 
problems in the base theory. Specialized provers and standard 
SMT solvers can be used for testing the satisfiability of the formulae 
obtained after the reduction. For a certain type of theory extension 
(namely for {\em local theory extensions}) this 
hierarchical reduction is sound and complete and -- 
if the formulae obtained this way belong to a fragment decidable in 
the base theory -- H-PILoT provides a decision procedure for testing 
satisfiability of ground formulae, and can also 
be used for model generation. 
\end{abstract}

\bigskip
\bigskip
\noindent {\Large\bf Table of Contents}

\bigskip

\bigskip

\contentsline {section}{\numberline {1}Introduction}{2}
\contentsline {section}{\numberline {2}Theoretical background}{3}
\contentsline {subsection}{\numberline {2.1}Local theory extensions}{3}
\contentsline {subsection}{\numberline {2.2}Examples of local extensions}{4}
\contentsline {section}{\numberline {3}Implementation}{5}
\contentsline {subsection}{\numberline {3.1}Generalities}{5}
\contentsline {subsection}{\numberline {3.2}Structure of the program}{6}
\contentsline {section}{\numberline {4}Modules of  {H-PILoT} }{7}
\contentsline {subsection}{\numberline {4.1}Preprocessing}{7}
\contentsline {subsection}{\numberline {4.2}Main algorithm}{7}
\contentsline {subsection}{\numberline {4.3}Post-processing}{8}
\contentsline {section}{\numberline {5}The input grammar}{9}
\contentsline {subsection}{\numberline {5.1}Declarations}{9}
\contentsline {subsection}{\numberline {5.2}Axiomatizations}{10}
\contentsline {section}{\numberline {6}Parameters of  {H-PILoT} }{13}
\contentsline {section}{\numberline {7}Error handling}{14}
\contentsline {section}{\numberline {8}Application areas}{14}
\contentsline {section}{\numberline {9}Examples}{16}
\contentsline {subsection}{\numberline {9.1}Monotone functions}{16}
\contentsline {subsection}{\numberline {9.2}Arrays}{18}
\contentsline {section}{\numberline {10}Example: Specifying the type information}{21}
\contentsline {subsection}{\numberline {10.1}Global constraints}{21}
\contentsline {subsection}{\numberline {10.2}Using standard types}{22}
\contentsline {section}{\numberline {11}Example: Handling data structures}{23}
\contentsline {subsection}{\numberline {11.1}Arrays}{23}
\contentsline {subsection}{\numberline {11.2}Pointers}{26}
\contentsline {section}{\numberline {12}Example: Using the built-in CNF translator}{28}
\contentsline {section}{\numberline {13}Extended locality}{29}
\contentsline {section}{\numberline {14}System evaluation}{31}
\contentsline {subsection}{\numberline {14.1}Test runs for H-PILoT}{31}
\contentsline {subsection}{\numberline {14.2}Test results}{32}
\contentsline {section}{\numberline {15}Examples}{33}
\contentsline {subsection}{\numberline {15.1}A case study}{33}
\contentsline {subsection}{\numberline {15.2}A run example of H-PILoT}{34}
\contentsline {subsection}{\numberline {15.3}Model generation and visualization}{38}
\contentsfinish



\section{Introduction}


H-PILoT (Hierarchical Proving by Instantiation in Local Theory extensions) 
is an implementation of the method for hierarchical reasoning 
in local theory extensions presented in 
\cite{GSW-ijcar-04,GSW-iandc-06,sofronie-cade05,sofronie-frocos07}:  it reduces the task of checking the 
satisfiability of a (ground) formula over the extension of a theory with 
additional function symbols subject to certain 
axioms (a set of clauses) to 
the task of checking the satisfiability of a formula over the base theory.
The idea is to replace the set of clauses which axiomatize the 
properties of the extension functions by a finite 
set of instances thereof. This reduction is polynomial in the size of 
the initial set of clauses and is always sound. It is complete 
in the case of so-called \emph{local extensions} \cite{sofronie-cade05}; 
in this case, 
it provides a decision procedure for validity for the universal fragment of 
the theory extension (or alternatively for satisfiability of ground clauses 
w.r.t.\ the theory extension) if the clauses obtained by the hierarchical 
reduction belong to a fragment for which satisfiability is 
decidable in the base theory. 
The satisfiability of the reduced set of clauses is then checked with 
a specialized prover for the base theory. 

State of the art SMT provers such as {\sf CVC3} \cite{cvc}, 
{\sf Yices} \cite{yicesa,yicesb} and {\sf Z3} \cite{z3,bjorner-z3^10} 
are very efficient for testing the satisfiability of {\em ground formulae} 
over standard theories, but use heuristics in the presence of 
{\em universally quantified} formulae, hence cannot detect 
{\em satisfiability} of such formulae.
H-PILoT recognizes a class of local 
axiomatizations, performs the instantiation and hands in a 
ground problem to the SMT provers or other specialized provers, 
for which they are know to terminate with a yes/no answer, so  
it can be used as a tool for 
steering standard SMT provers, in order to provide decision procedures 
in the case of local theory extensions. 
H-PILoT can also be used for generating models of satisfiable formulae; 
and even more, it can be coupled to programs with graphic facilities  
to provide graphical representations of these models. 
Being a decision procedure for many theories important in verification, 
H-PILoT is extremely helpful for 
deciding truth or satisfiability in a large variety of verification 
problems.

\medskip
\noindent This is an extended version of the description of H-PILoT presented at 
CADE 22 \cite{sofronie-ihlemann:hpilot}. 

\section{Theoretical background}
\label{theory}

Many problems in mathematics and computer science can be reduced to proving
the satisfiability of conjunctions of literals in a background theory 
(which can be the extension of a base theory with additional functions -- 
e.g., free, monotone, or recursively defined -- or a combination of theories). 
Considerable work has been dedicated to the task of identifying situations 
where reasoning in extensions and combinations of theories can be done 
efficiently and accurately. The most important issues which need to be 
addressed in this context are:
\begin{itemize} 
\item[(i)] finding possibilities of reducing the search space without 
losing completeness, and 
\item[(ii)] making modular or hierarchical reasoning possible.
\end{itemize}
In \cite{GivanMcAllester92,mcallester,McAllester-acm-tocl-02}, 
Givan and McAllester
introduced and studied the so-called ``local inference systems'', for which
validity of ground Horn clauses can be checked in polynomial time.  A
link between this proof theoretic notion of locality and algebraic
arguments used for identifying classes of algebras with a word problem
decidable in PTIME \cite{Burris95} was established in
\cite{ganzinger:local_reasoning}.  
In \cite{GSW-ijcar-04,GSW-iandc-06,sofronie-cade05}
these results were further extended to so-called {\em local
  extensions} of theories.
Locality phenomena were also studied in the verification 
literature, mainly motivated by the necessity of devising
methods for efficient reasoning in theories of pointer structures  
\cite{necula-mcpeak} and arrays \cite{bradley-manna}. 

In \cite{sofronie-ihlemann-jacobs:tacas}
we showed that these results are instances of a general
concept of locality of a theory extension -- parameterized by a 
closure operator on ground terms. 
The main idea of locality and local extensions is to limit the search space for counterexamples (hence the name).

\subsection{Local theory extensions}
We consider the following setup.
Let ${\cal T}_0$ be a theory in some signature $\Sigma_0$.
We consider extensions 
${\cal T}_1 = {\mathcal T}_0 \cup {\mathcal K}$ of ${\cal T}_0$ with  function symbols in a set $\Sigma_1$ 
(extension functions) whose properties are axiomatized 
by a set ${\mathcal K}$ of (universally closed) $\Sigma_0 \cup \Sigma_1$-clauses. Let  
$\Sigma_c$ be an additional set of constants.

\bigskip
\noindent {\bf Task.} Let $G$ be a set of 
ground $\Sigma_0 \cup \Sigma_1 \cup \Sigma_c$-clauses. 
We want to check whether or not $G$ is satisfiable  
w.r.t.\ ${\cal T}_0 \cup {\mathcal K}$. 

\bigskip
\noindent {\bf Method.} Let ${\mathcal K}[G]$ be the set of those instances of ${\mathcal K}$ 
in which every subterm starting with an extension function
is a ground subterm already appearing in ${\mathcal K}$ or $G$.
If $G$ is unsatisfiable w.r.t.\ ${\cal T}_0 \cup {\mathcal K}[G]$ then it is also 
unsatisfiable w.r.t.\ ${\cal T}_0 \cup {\cal K}$. The converse is not necessarily 
true.
\begin{definition}
We say that the extension ${\cal T}_0 \cup {\mathcal K}$ of ${\cal T}_0$
is \emph{local}  if it satisfies the following 
condition\footnote{It is easy to check that the formulation we 
give here and that in \cite{sofronie-cade05} are equivalent.}:  
\begin{tabbing}
\hspace{1ex} \= ${\sf (Loc})$ \qquad \= For every set $G$ of ground $\Sigma_0 \cup \Sigma_1 \cup \Sigma_c$-clauses it holds that\\
\>\> $\T_0 \cup {\cal K} \cup G \models \bot$ if and only if $\T_0 \cup \K[G] \cup G \models \bot.$
\end{tabbing}
\end{definition}
Thus, the method is sound and complete for {\em local theory extensions}.

\begin{theorem}[\cite{sofronie-cade05}]
Assume that the extension 
${\mathcal T}_0 \subseteq {\mathcal T}_1 = {\mathcal T}_0 \cup {\mathcal K}$ is local
and let $G$ be a set of ground clauses. 
Let ${\mathcal K}^0 \cup G^0 \cup D$ be the purified form
of ${\mathcal K} \cup G$ obtained by  introducing fresh constants
for the $\Sigma_1$-terms, adding their definitions $d \approx f(t)$ to $D$, 
and replacing $f(t)$ in $G$ and ${\mathcal K}[G]$ by $d$.
(Then $\Sigma_1$-functions occur only in $D$ in unit clauses of the form $d \approx f(t)$.) 
The following are equivalent.
\begin{enumerate}
\item ${\mathcal T}_0 \cup {\mathcal K} \cup G$ has a (total) model.
\item ${\mathcal T}_0 \cup {\mathcal K}[G] \cup G$ has a partial model where all subterms of ${\mathcal K}$ and $G$ 
and all $\Sigma_0$-functions are defined.
\item ${\mathcal T}_0 \cup {\mathcal K}^0 \cup G^0 \cup {\sf Con}^0$ has a total model, where \\ ${\sf Con}^0 := \{\, \bigwedge_{i=1}^n c_i \approx d_i  \rightarrow c \approx d \,|\;
            f(c_1,...,c_n) \approx c,  f(d_1,...,d_n) \approx d \in D \}$.
\end{enumerate}
\label{th-reduction}
\end{theorem}
A variant of this notion, namely $\Psi$-locality,  
was also studied, where the set of instances to be taken into account 
is ${\cal K}[\Psi(G)]$, where $\Psi$ is a closure operator which may 
add a (finite) number of new terms to the subterms of $G$.
We also analyzed a generalized version of locality, in which the 
clauses in ${\cal K}$ and the set $G$ of ground clauses are allowed 
to contain first-order $\Sigma_0$-formulae.  

\subsection{Examples of local theory extensions} 
Among the theory extensions which we proved to be  local or $\Psi$-local in previous work 
are: 
\begin{itemize}
\item a fragment of the theory of pointers with stored scalar information in the nodes introduced in \cite{necula-mcpeak}, further analyzed in 
\cite{sofronie-ihlemann-jacobs:tacas,faber-ihlemann-jacobs-sofronie-ifm10}; 
\item a fragment of the theory of arrays with integer indices, and elements in a given 
theory introduced in \cite{bradley-manna}, further analyzed in 
\cite{sofronie-ihlemann-jacobs:tacas};  
\item theories of functions over an ordered domain
or over a numerical domain satisfying  
monotonicity or boundedness conditions \cite{sofronie-cade05,viorica:interpolation,sofronie-ihlemann-ismvl-07}; 
\item various combinations of such extensions \cite{sofronie-frocos07,sofronie-ihlemann-jacobs:tacas}.
\end{itemize}
We can also consider successive extensions of theories: ${\cal T}_0 \subseteq 
{\cal T}_0 \cup {\cal K}_1 \subseteq \dots \subseteq {\cal T}_0 \cup {\cal K}_1 \cup \dots \cup {\cal K}_n$. 
If every 
variable in ${\cal K}_i$ occurs below a function symbol in $\Sigma_i$, 
this reduction process 
can be iterated \cite{sofronie-ihlemann-jacobs:tacas}. 

For local theory extensions, Theorem~\ref{th-reduction} allows us to
reduce the original problem to a satisfiability problem over the base
theory $\T_0$.

\section{Implementation}
\label{ch-hpilot}

\vspace{-2mm}
The software system \hpilot (Hierarchical
Proving by Instantiation in Local Theory extensions) for hierarchical
reasoning in local theory extensions is implemented as follows: 
A given proof task (set of
ground clauses), over the extension of a theory with functions
axiomatized by a set of clauses, is reduced to an equi-satisfiable
ground problem over the base theory in the manner of
Theorem~\ref{th-reduction}.
After \hpilot has carried out this reduction, it hands over the
transformed problem to a dedicated prover for the base theory.  This
reduction is always sound.  For local theory extensions the
hierarchical reduction is sound and complete.  If the formulas
obtained in this way belong to a fragment decidable in the base
theory, H-PILoT provides a decision procedure for testing
satisfiability of ground formulas.  If the reduced formulas are
satisfiable (modulo the base theory), \hpilot can be used for model
generation, which is of great help in detecting and localizing errors.

\vspace{-2mm}
\subsection{Generalities}

\vspace{-2mm}
H-PILoT is implemented in
Ocaml\footnote{\url{http://caml.inria.fr/ocaml/index.en.html}}.  The
system, together with a manual and examples, can be downloaded from
\url{www.mpi-inf.mpg.de/~ihlemann/software/}.  There is both a 32-bit
and a 64-bit Linux version available. 

To improve user-friendliness, a clausifier has
also been integrated into \hpilot (Sect.~\ref{sec-clausification}).
H-PILoT recognizes a class of local axiomatizations; 
it has advanced abilities to handle the common data
structures of \emph{arrays} (Sect.~\ref{sec-arrays}) and
\emph{pointers} (Sect.~\ref{sec-pointers})\footnote{\hpilot automatically 
detects whether a given specification falls within the local fragment 
of these theories.}.  
H-PILoT performs the
instantiation and hands in a ground problem to the SMT provers or
other specialized provers, for which they are known to terminate with
a yes/no answer, so it can be used as a tool for steering standard SMT
provers, in order to provide decision procedures in the case of local
extensions.
The provers integrated with \hpilot are the general-purpose prover SPASS (\cite{spass});
the SMT-solvers Yices (\cite{yicesa,yicesb}), CVC3 (\cite{cvc}) and Z3 (\cite{z3});
and the prover Redlog (\cite{redlog}) for non-linear real problems.
State-of-the-art SMT provers, such as the ones above, are very
efficient for testing the satisfiability of ground formulas over
standard theories, such as linear arithmetic (real, rational or
integer), but use heuristics in the presence of universally quantified
formulas, hence, cannot reliably detect satisfiability of such
formulas.  However, if SMT solvers are used for finding software bugs,
being able to detect the actual satisfiability of satisfiable
sets of formulas is crucial
(cf.~\cite{smt@microsoft,z3,DBLP:conf/cav/GeM09,bjorner-z3^10}).
For local theory extensions, \hpilot offers the possibility of detecting 
satisfiability and of constructing models for satisfiable sets of clauses.
On request, \hpilot provides an extensive step-by-step trace of the
reduction process, making its results verifiable
(Sect.~\ref{sec-run}).
 
\hpilot 
has been used in large case studies, where its 
ability (1) to handle \emph{chains} of extensions, (2) to detect 
unsatisfiability {\em and} satisfiability, and (3) to construct models 
of satisfiable sets of clauses has been crucial. 

\subsection{Structure of the program}

\begin{wrapfigure}[30]{r}{.4\textwidth}
 \centering 
\vspace{-12mm}
  \includegraphics[width=0.38\textwidth]{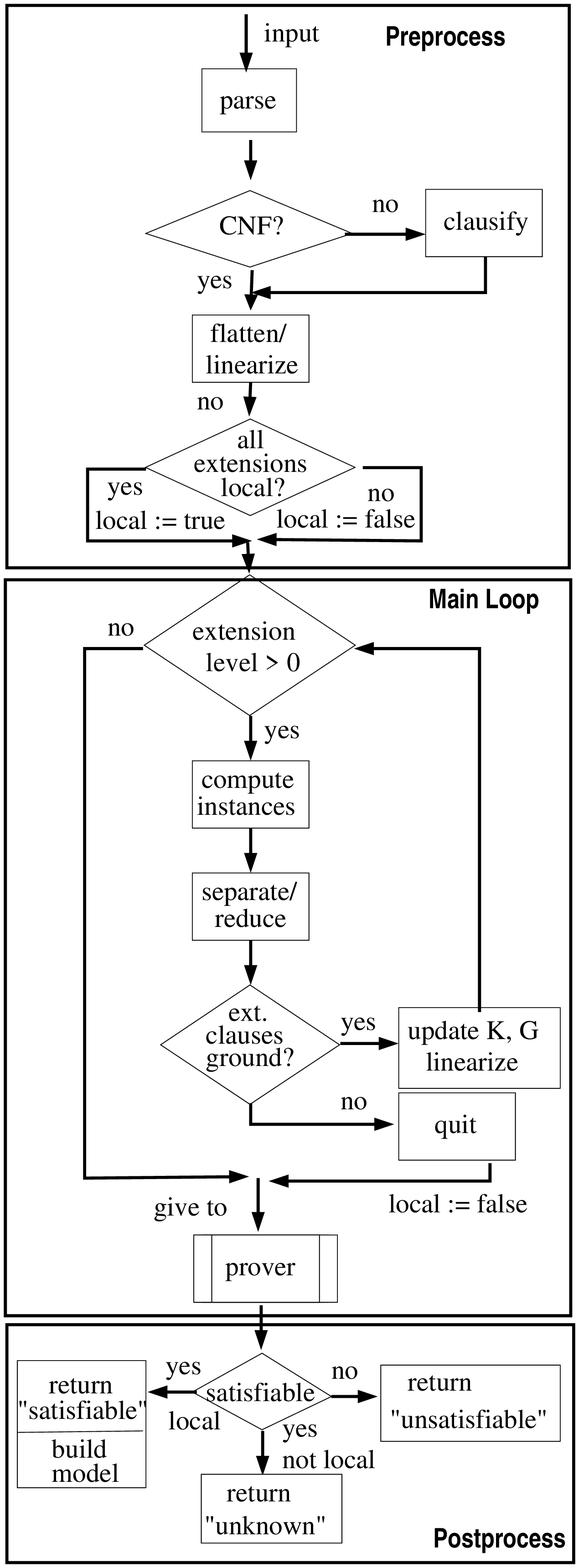}
  \caption{H-PILoT Structure}
  \label{fig:struct}
\end{wrapfigure}

The main algorithm which hierarchically reduces a decision problem in a theory extension to a decision problem  in the base
theory can be divided into a preprocessing part,
the main loop and a post-processing part; see Figure \ref{fig:struct}.

\medskip
\noindent {\bf Preprocessing.} 
The input is read and parsed. If it is detected to be in SMT format,
we set the options to ``use arithmetic'' (e.g., $+$, $-$,... are
predefined).  If the input is not in clause normal form (CNF), it is
translated to CNF, then the input is flattened and linearized.  The
program then checks if the clauses in the axiomatization given are
local extensions and sets the flag \verb+-local+ to true/false. This
ends the preprocessing phase.

\medskip
\noindent {\bf Main algorithm.}
The main loop proceeds as follows: We consider chains of extensions
$\T_0 \subseteq \T_1 \subseteq \cdots \subseteq \T_n$, where $\T_i = \T_0
\cup \bigcup_{j = 1}^i \K_j$ of $\T_0$ with function symbols in a set
$\Sigma_i$ (extension functions) whose properties are axiomatized by a
set $\K_i$ of $(\Pi_0  \cup \bigcup_{j = 1}^i \Sigma_j)$-clauses. 

Let $i = n$.  As long as the extension level $i$ is greater than $0$,
we compute $\K_i[G]$ ($\K_i[\Psi]$ for arrays). If no separation of
the extension symbols is required, we stop here (the result will be
passed to an external prover that can reason about the extension of 
the theory $\T_0$
with free function symbols in $\Sigma_n$). 
Otherwise, we perform the hierarchical
reduction by purifying $\K_i$ and $G$ (to $\K_i^0$, $G^0$
respectively)
 and by adding corresponding instances of the congruence
axioms ${\sf Con}_i$. To prepare for the next iteration, we transform
the clauses into the form $\forall \x. \Phi \vee \K_i$ (compute prenex
form, skolemize). If $\K_i[G]$ ($i > 1$) is not ground, we quit
with a corresponding message. Otherwise we set $G' := \K_i^0 \wedge
G^0 \wedge {\sf Con}_i$ and $\T' := \T \setminus \{\K_i\}$. We flatten
and linearize $\K'$ and decrease $i$. If level $i = 0$ is reached
$G'$ is handed to an external prover.

\medskip
\noindent {\bf Post-processing.}
If the answer is ``unsatisfiable'' then $G \models_{{\T}_n} \falsum$. If
the answer is ``satisfiable'' and all extensions were local, then $G$
is satisfiable w.r.t. $\T_n$ and we know how to build a model. If the
answer is satisfiable but we do not know that all extensions are
local, or if the instantiated clauses (of level $> 1$) were not
ground, we answer ``unknown''.

\section{Modules of \hpilot}

We present the different parts of \hpilot in more detail.

\subsection{Preprocessing}

H-PILoT receives as input a many-sorted specification of the
signature; a specification of the hierarchy of local extensions to be
considered; an axiomatization $\K$ of the theory extension(s); a
set $G$ of ground clauses containing possibly additional constants.
H-PILoT allows the following preprocessing functionality.

\medskip

\noindent {\bf Translation to clause form.} H-PILoT provides a
translator to clause normal form (CNF) for ease of use.  First-order
formulas can be given as input; \hpilot translates them into CNF. In
the present implementation, the CNF translator does not provide the
full functionality of FLOTTER (\cite{nonnengart:cnf}) -- it has only
restricted subformula renaming -- but is powerful enough for most
applications.

\medskip

\noindent
{\bf Flattening/linearization.} 
Methods for recognizing local theory extensions usually require that
the clauses in the set ${\mathcal K}$ extending the base theory are
\emph{flat} and \emph{linear}, which does nothing to improve
readability.  If the flags {\sf -linearize} and/or {\sf -flatten} are
used then the input is flattened and/or linearized (the general
purpose flag {\sf -preprocess} may also be used).  H-PILoT allows the
user to enter a more intelligible non-flattened version and will
perform the flattening and linearization of ${\mathcal K}$.

\medskip

\noindent
{\bf Recognizing syntactic criteria which imply locality.} 
\noindent Examples of local extensions include (fragments of) the
theories of the common data structures: the theory of arrays (see
Section~\ref{sec-arrays}) and the theory of pointers (see
Section~\ref{sec-pointers}), respectively (and also iterations and
combinations thereof).  In the preprocessing phase H-PILoT analyzes
the input clauses to check whether they are in one of these fragments.

\begin{itemize}
\item If the flag {\sf -array} is on, \hpilot checks if the input is
  in the ``array property fragment''.
\item If the keyword ``pointer'' is detected, \hpilot checks if the
  input is in the appropriate pointer fragment and adds missing
  ``nullability'' terms, i.e., it adds premises of the form ``$t \neq
  {\sf null}$'', in order to relieve the user of this clerical labor.
\end{itemize}

\noindent If the answer is ``yes'' then we
know that the extensions we consider are local, i.e., that H-PILoT can
be used as a decision procedure.  

\subsection{Main algorithm}

The main algorithm hierarchically reduces a decision problem in a theory extension to a decision problem in the base theory.

\medskip
Given a set of clauses ${\mathcal K}$ and a set of ground clauses $G$,
the algorithm we use carries out a hierarchical reduction of $G$ to a
set of formulas in the base theory.  It then hands over the new
problem to a dedicated prover such as Yices, CVC3 or Z3.  H-PILoT is
also coupled with Redlog (for handling non-linear real arithmetic) and
with SPASS\footnote{H-PILoT only calls one of these solvers once.}.

\begin{wrapfigure}[24]{r}{.4\textwidth}
  \centering 
\vspace{-4mm}
  \includegraphics[height=10cm,width=.38\textwidth]{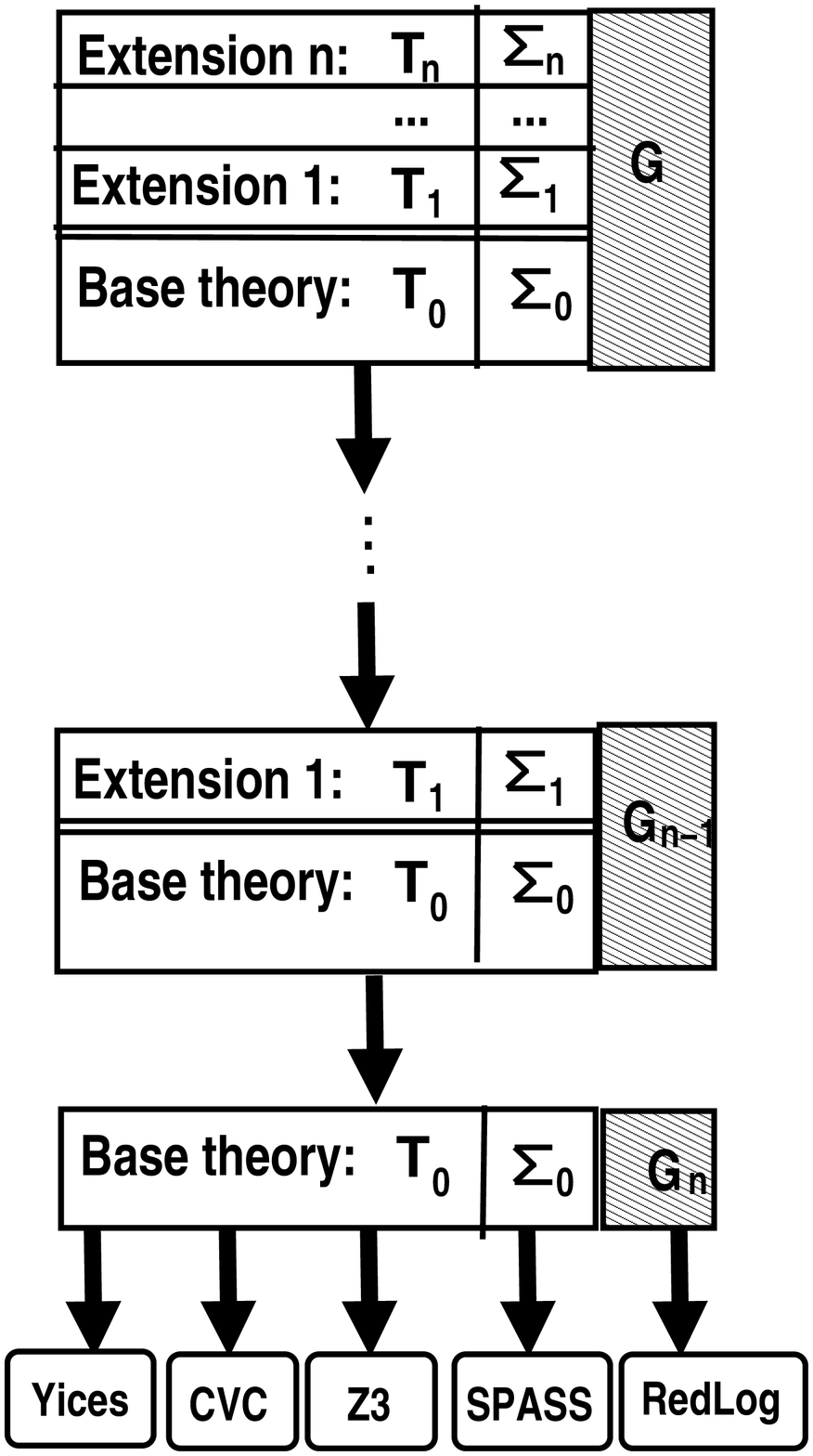}
  \label{fig:hierarchy}
\end{wrapfigure}

\medskip
\noindent
{\bf Loop.} 
For a chain of local extensions:  
\begin{eqnarray*}
\T_0 & \subseteq &  \T_1  =  \T_0 \cup \K_1  \subseteq  \T_2 = \T_0  \cup \K_1 \cup \K_2 \\
           & \subseteq & ... \subseteq  \T_n = \T_0 \cup \K_1 \cup...\cup \K_n.
\end{eqnarray*}
a satisfiability check w.r.t.\ the last extension can be reduced (in
$n$ steps) to a satisfiability check w.r.t.\ $\T_0$.  The only caveat
is that at each step the reduced clauses $\K_{i}^0 \cup G^0 \cup {\sf
  Con}^0$ need to be ground.  Groundness is assured if each variable
in a clause appears at least once under an extension function.  In
that case, we know that at each reduction step the total clause size
only grows polynomially in the size of $G$
(\cite{sofronie-cade05}).  H-PILoT ~allows the user to specify a
chain of extensions by tagging the extension functions with their
place in the chain (e.g., if $f$ belongs to $\K_3$ but not to $\K_1
\cup \K_2$ it is declared as level 3).

Let $i = n$. As long as the extension level $i$ is larger than $0$, we
compute $\K_i[G]$ ($\K_i[\Psi(G)]$ in case of arrays).  If no
separation of the extension symbols is required, we stop here (the
result will be passed to an external prover). Otherwise, we perform
a hierarchical reduction  by purifying
$\K_i$ and $G$ (to $\K_i^0, G^0$ respectively) and by adding
corresponding instances of the congruence axioms ${\sf Con}_i$.  To
prepare for the next iteration, we transform the clauses into the form
$\forall x. \Phi \vee \K_{i}$ (compute prenex form, skolemize). If
$\K_i[G] / \K_i^0$ is not ground, we quit with a corresponding
message. Otherwise our new proof task $G'$ becomes $G' := \K_i^0
\wedge G^0 \wedge {\sf Con}_i$, our new extension clauses are $\K' :=
\K_{i-1}$ and our new base theory becomes $\T' := \T_{i-1} \setminus
\{\K_{i-1}\}$. We flatten and linearize $\K'$ and decrease $i$.  If
level $i = 0$ is reached, we exit the main loop and $G'$ is handed to
an external prover.  Completeness is guaranteed if all extensions are
known to be local and if each reduction step produces a set of ground
clauses for the next step.

\subsection{Post-processing}

Depending on the answer of the external provers to the satisfiability 
problem $G_n$, we can infer whether 
the initial set $G$ of clauses was satisfiable or not. 

\begin{itemize}
\item If $G_n$ is unsatisfiable w.r.t.\ $\T_0$ then we know that $G$ is unsatisfiable. 

\item If $G_n$ is satisfiable, but H-PILoT failed to detect this, and the
  user did not assert the locality of the sets of clauses used in the
  axiomatization, its answer is ``unknown''.

\item If $G_n$ is satisfiable and H-PILoT detected the locality of the
  axiomatization, then the answer is ``satisfiable''.  In this case,
  H-PILoT takes advantage of the ability of SMT-solvers to provide
  counterexamples for the satisfiable set $G_n$ of ground clauses and
  specifies a counterexample of $G$ by translating the basic
  SMT-model of the reduced proof task to a model of the original
  proof task. This improves readability greatly, especially when we have a
  chain of extensions.  The counterexamples can be graphically
  displayed using Mathematica (cf.   Section~\ref{sec-model-generation}). 
 This is currently done separately; an integration with Mathematica is 
 planned for the future.
\end{itemize}

\section{The input grammar}

\newcommand{\nt}[1]{$\langle${\it #1}$\rangle$}
\renewcommand{\t}[1]{{\tt #1}}
\newcommand{\e}{$\epsilon\;$}
\newcommand{\m}{$\mid\:$}

The input file consists of a {\em declaration} part (for function symbols 
in the base theory, for extension functions, for relations, and constants), 
specifications of the {\em types} and of the base theory, 
a part containing {\em axiomatizations} of base theory/extension functions; 
and a part containing the set of {\em  ground clauses} whose 
satisfiability is being checked.    

\setlength{\parindent}{0cm}

\medskip

\nt{start} ::= \nt{base\_functions} \nt{extension\_functions} \nt{relations} \nt{constants} \nt{interval} \\
               \hspace*{12ex}\nt{baseTheory} \nt{formulasOrClauses} \nt{groundformulas} \nt{query}

\subsection{Declarations}

{\bf Type declarations:} 
We allow for declarations of standard types:

\medskip
\begin{tabbing}
\nt{domain} \= ::= \t{bool} 
  \m \t{int} 
  \m \t{real}
  \m \t{pointer}
  \m \t{pointer\#} \nt{int}\\
  \> \m \t{scalar}
  \m \t{free}
  \m \t{free\#} \nt{int}
\end{tabbing}

\medskip
\noindent Declarations of simple types such as intervals are also allowed:

\begin{tabbing}
\nt{interval}  \= ::=  $\;$\e\\
  \>\m \t{Interval} \t{:=} \nt{int} \nt{sm} \nt{identifier}\t{;}\\
  \>\m \t{Interval} \t{:=} \nt{identifier} \nt{sm} \nt{int}\t{;}\\
  \>\m \t{Interval} \t{:=} \nt{int} \nt{sm} \nt{identifier} \nt{sm} \nt{int}\t{;}
\end{tabbing}

\nt{int} ::= {\it any non-negative number.}

\medskip

\nt{sm} ::= \t{<=}  \m  \t{<}

\medskip

Further details are given in Section~\ref{types}. 

\medskip
\noindent {\bf Function and relation declarations.}
The declaration part contains sort and type declarations for the 
function symbols in the base theory, for the extension function symbols, 
for the relation symbols and for the constants:

\medskip

\nt{base\_functions} ::= \t{Base\_functions} \t{:=} \t{\{} \nt{function\_list} \t{\}}

\smallskip

\nt{extension\_functions} ::= \t{Extension\_functions} \t{:=} \t{\{} \nt{function\_list} \t{\}} 

\smallskip

\nt{relations} ::= \t{Relations} \t{:=} \t{\{} \nt{relation\_list} \t{\}}

\smallskip

\nt{constants} ::= \e  \m \t{Constants} \t{:=} \t{\{} constant\_list \t{\}}

\bigskip

\nt{constant\_list} ::= \nt{constant} \nt{additional\_constants}

\smallskip

\nt{additional\_constants} ::= \e  \m \t{,} \nt{constant} \nt{additional\_constant}

\begin{tabbing}
\nt{constant} \= ::= 
   \t{(} \nt{identifier} \t{,} \t{bool} \t{)}\\
  \>\m \t{(} \nt{identifier} \t{,} \t{int}  \t{)}\\
  \>\m \t{(} \nt{identifier} \t{,} \t{real} \t{)}\\
  \>\m \t{(} \nt{identifier} \t{,} \t{scalar} \t{)} \\
  \>\m \t{(} \nt{identifier} \t{,} \t{pointer} \t{)}\\
  \>\m \t{(} \nt{identifier} \t{,} \t{pointer\#} \nt{int} \t{)}\\
  \>\m \t{(} \nt{identifier} \t{,} \t{free} \t{)}\\
  \>\m \t{(} \nt{identifier} \t{,} \t{free\#} \nt{int} \t{)}
\end{tabbing}

\bigskip

\nt{function\_list} ::= \e \m \nt{function} \nt{additional\_functions}

\smallskip

\nt{additional\_functions} ::= \e \m \t{,} \nt{function} \nt{additional\_functions}

\smallskip

\nt{relation\_list} ::= \e \m \nt{relation} \nt{additional\_relations}

\smallskip

\nt{additional\_relations} ::= \e \m \t{,} \nt{relation} \nt{additional\_relations}

\medskip

\medskip
We allow for predefined relation declarations (e.g.\ for relations
such as $\leq$ or $<$ as well as for new relation declarations. 
In each case we specify together with the relation symbols also their 
arity. 

\medskip
\nt{relation} ::= \t{(} \nt{uneqs} \t{,} \nt{int} \t{)}
  \m  \t{(} \nt{identifier} \t{,} \nt{int} \t{)}

\medskip
We allow for several forms of function declaration: We can declare 
the number of arguments of the function ((1),(2)); 
the number of arguments and 
the level (for predefined arithmetical operations over the integers (3), 
or reals (4) or for uninterpreted functions (5)); the number of arguments,  
the level, and the sort of the domain and codomain (without repetitions 
if the domain and codomain are the same (6); separately specified if 
they are different (7)). 

\begin{tabbing}
\nt{function} \= ::= \t{(} \nt{identifier} \t{,} \nt{int} \t{)} ~~~~~~~~~~~~~~~~~~~~~~~~~~~~~~~~~~~~~~~~~~~~~ \=  (1)\\
  \>\m \t{(} \nt{arithop} \t{,} \nt{int} \t{)} \> (2)\\
  \>\m \t{(} \nt{arithop} \t{,} \nt{int} \t{,} \nt{int} \t{,} \t{int} \t{)}  \> (3)\\
  \>\m \t{(} \nt{arithop} \t{,} \nt{int} \t{,} \nt{int} \t{,} \t{real} \t{)} \>  (4)\\
  \>\m \t{(} \nt{identifier} \t{,} \nt{int} \t{,} \nt{int} \t{)} \> (5)\\
  \>\m \t{(} \nt{identifier} \t{,} \nt{int} \t{,} \nt{int} \t{,} \nt{domain} \t{)} \> (6) \\
  \>\m \t{(} \nt{identifier} \t{,} \nt{int} \t{,} \nt{int} \t{,} \nt{domain} \t{,} \nt{domain} \t{)} \> (7)
\end{tabbing}
At the moment declarations of functions which accept arguments of different 
sorts are not supported.

\subsection{Axiomatizations}

We support axiomatizations for the base theory:

\medskip
\nt{base\_theory} ::= \e  \m \t{Base} \t{:=} \nt{clause\_list}

\medskip

\noindent 
axiomatizations of the properties of the extension functions: 
\begin{tabbing}
\nt{formulasOrClauses} ::= \e
  \=\m \nt{formulas}
  \m \nt{clauses}\\
  \>\m \nt{formulas}\nt{clauses}
  \m \nt{clauses}\nt{formulas}
\end{tabbing}
as well as input of the ground formulae whose (un)satisfiability 
is being checked:

\medskip

\nt{formulas} ::= \t{Formulas} \t{:=} \nt{formula\_list}

\smallskip

\nt{formula\_list} ::= \nt{formula} \m \nt{formula} \t{;} \nt{additional\_formulas}
 
\smallskip

\nt{additional\_formulas} ::= \e  \m \nt{formula} \t{;} \nt{additional\_formulas}

\begin{tabbing}
\nt{formula} \= ::= \nt{atom}\\
  \>\m \t{NOT} \t{(} \nt{formula} \t{)}\\
  \>\m \t{OR}  \t{(} \nt{formula} \nt{formula\_plus} \t{)}\\
  \>\m \t{AND} \t{(} \nt{formula} \nt{formula\_plus} \t{)}\\
  \>\m \t{(} \nt{formula} \t{-->} \nt{formula} \t{)} \\
  \>\m \t{(} \nt{formula} \t{<-->} \nt{formula} \t{)} \\
  \>\m \t{(} \t{FORALL} \nt{variables} \t{)} \t{.} \nt{formula} \\
  \>\m \t{(} \t{EXISTS} \nt{variables} \t{)} \t{.} \nt{formula} 
\end{tabbing}

\nt{formula\_plus} ::= \t{,} \nt{formula} \nt{formula\_star}

\smallskip

\nt{formula\_star} ::= \e
  \m \t{,} \nt{formula} \nt{formula\_star}

\smallskip

\nt{ground\_formulas} ::= \e
  \m \t{Ground\_Formulas} \t{:=} \nt{formula\_list}

\bigskip
\nt{query} ::= \t{Query} \t{:=} \nt{ground\_clauses}

\bigskip

\nt{clauses} ::=  \t{Clauses} \t{:=} \nt{clause\_list}

\smallskip

\nt{base\_clause\_list} ::= \e \m \nt{clause} \t{;} \nt{additional\_base\_clauses}

\smallskip

\nt{additional\_base\_clauses} ::= \e \m \nt{base\_clause} \t{;} \nt{additional\_base\_clauses}

\smallskip

\nt{base\_clause} ::= \nt{clausematrix} 
  \m \nt{universalQuantifier} \nt{clausematrix}

\smallskip

\nt{clause\_list} ::= \nt{clause} \m \nt{clause} \t{;} \nt{additional\_clauses}

\smallskip

\nt{additional\_clauses} ::= \e
  \m \nt{clause} \t{;} \nt{additional\_clauses}

\begin{tabbing}
\nt{clause} \= ::=  \nt{clausematrix} \\
  \>\m \nt{universalQuantifier} \nt{clausematrix} \\
  \>\m \t{\{} \nt{formula} \t{\}} \t{OR} \nt{clausematrix}\\
  \>\m \nt{universalQuantifier} \t{\{} \nt{formula} \t{\}} \t{OR} \nt{clausematrix}\\
  \>\m \t{\{} \nt{formula} \t{\}} \t{-->} \nt{clausematrix}\\
  \>\m \nt{universalQuantifier} \t{\{} \nt{formula} \t{\}} \t{-->} \nt{clausematrix}
\end{tabbing}

\nt{universalQuantifier} ::= \t{(} \t{FORALL} \nt{variables} \t{)} \t{.}
 
\smallskip

\nt{variables} ::= \nt{name} \nt{additional\_variable}

\smallskip

\nt{additional\_variables} ::= \e \m \t{,} \nt{name} \nt{additional\_variables}

\smallskip

\nt{ground\_clauses} ::= \e
  \m \nt{clausematrix} \t{;} \nt{ground\_clauses}

\smallskip

\nt{clausematrix} ::=  \nt{literal} \m \nt{disjunctive\_clause} \m \nt{sorted\_clause}

\bigskip

\nt{disjunctive\_clause} ::= \t{OR} \t{(} \nt{literal} \nt{literal\_plus} \t{)}

\smallskip
\nt{literal\_plus} ::= \t{,} \nt{literal} \nt{literal\_star}

\smallskip

\nt{literal\_star} ::= \e
  \m \t{,} \nt{literal} \nt{literal\_star}

\bigskip

\nt{sorted\_clause} ::= \nt{atom\_list} \t{-->} \nt{atom\_list}

\smallskip

\nt{atom\_list} ::= \e  \m \nt{atom} \nt{atom\_star}

\smallskip

\nt{atom\_star} ::= \e \m \t{,} \nt{atom}  \nt{atom\_star}

\bigskip

\nt{literal} ::= \nt{atom}  \m \t{NOT} \t{(} \nt{atom} \t{)}

\smallskip 

\nt{atom} ::= \nt{equality\_atom}
  \m \nt{ineq\_atom}
  \m \nt{predicate\_atom}

\smallskip

\nt{equality\_atom} ::=  \nt{term} \t{=} \nt{term}

\smallskip

\nt{ineq\_atom} ::= \nt{term} \nt{uneqs} \nt{term}

\smallskip
                                                           
\nt{predicate\_atom} ::= \nt{identifier} \t{[} \nt{term} \nt{additional\_terms} \t{]}

\smallskip

\nt{arguments} ::= \nt{term} \nt{additional\_terms}

\smallskip

\nt{additional\_terms} ::= \e  \m \t{,} \nt{term} \nt{additional\_terms}

\begin{tabbing}
\nt{term} \= ::= \nt{name}\\
  \>\m \nt{operator} \t{(} \nt{arguments} \t{)}\\
  \>\m \nt{array} \t{(}  \nt{arguments} \t{)}\\   
  \>\m \nt{update} \t{(} \nt{arguments} \t{)}\\  
  \>\m \nt{term\_arith} \nt{arithop} \nt{term\_arith}
\end{tabbing}

\begin{tabbing}
\nt{term\_arith} \= ::= \nt{name}  \\
   \> \m \nt{operator} \t{(} \nt{arguments} \t{)}\\
   \> \m \t{(} \nt{term\_arith} \nt{arithop} \nt{term\_arith} \t{)}
\end{tabbing}

\nt{arithop} ::= \t{+}  \m \t{-} \m \t{*}  \m \t{/}

\smallskip

\nt{uneqs} ::=  \t{<=} \m \t{>=} \m \t{<} \m \t{>} 

\smallskip

\nt{operator} ::= \nt{identifier}

\begin{tabbing}
\nt{array:} \= ::=   \t{write} \t{(} \nt{identifier} \t{,} \nt{term} \t{,} \nt{term} \t{)}\\
              \> \m \t{write} \t{(} \nt{array} \t{,} \nt{term} \t{,} \nt{term} \t{)}
\end{tabbing}

\begin{tabbing}
\nt{update} \= ::= \t{update} \t{(} \nt{identifier} \t{,} \nt{term} \t{,} \nt{term} \t{)}\\
  \>\m \t{update} \t{(} \nt{update} \t{,} \nt{term} \t{,} \nt{term} \t{)}
\end{tabbing}

\nt{name} ::= \nt{identifier}

\begin{tabbing}
\nt{identifier} ::= \= {\it any string consisting of letters and numbers starting with a letter.}\\
                    \> {\it It may end with ``\t{'}''.}
\end{tabbing}

\section{Parameters of \hpilot}
\label{parameters}

\hpilot has several input parameters controlling its behavior. They can be listed by calling
\verb+hpilot.opt -help+.
\bigskip

\noindent
\begin{tabular}{||ll||} \hline
  -min & Use minimal locality. Currently, this is only relevant \\
       & for the array property fragment.\\
  -prClauses & Produce output: print all the clauses calculated and used.\\
  -noProver &Do not hand over to prover, just produce output.\\
  -arith &Use arithmetic. 'plus','+','-' etc. are predefined.\\
         & Numerals (names for integers) must be used preceded by \\
& underscore $\_$, e.g., '\_3'.\\
  -yices &Use Yices as background solver: 'plus', '+' etc. \\
         &are predefined as are the order relations $\leq, \geq, <, >$.\\
         &Numbers can also be given in the input.\\
         &Numbers are integers by default\\ 
      &(use '-real' for real numbers).\\
  -cvc   &Use CVC as background solver.\\
         &Arithmetic is predefined as with '-yices'.\\
  -z3	 &Use Z3 as background solver.\\
         &Arithmetic is predefined as with '-yices'.\\
  -flatten &Flatten clauses first.\\
  -linearize &Linearize clauses first.\\
  -flattenQuery &Flattens the proof task first.\\
  -preprocess &Preprocess input: flatten/linearize clauses, flatten proof task.\\
              &In array-context: split clauses which contain inequalities\\
              & like $i \neq j$ into two clauses.\\
  -noSeparation &Stop at calculating the instances $\K[G]$. Don't introduce \\
                &names for extension terms and don't reduce to base theory.\\
  -unPseudofy &Eliminate pseudo-quantifiers like $\forall i. i=3...$ \\
              & before handing over to a prover.\footnotemark\\
  -noProcessing &No computation. Just translate into prover syntax and \\
                &hand over. Overrides '-preprocess'.\\
                &When using this flag one should provide the domains\\
                &of functions too. When used in combination with CVC \\
                &there may arise problems with boolean types.\footnotemark\\
  -clausification &Toggle clausification (true/false). Default is 'true'. \\
	 &'false' implies '-noProcessing'.\\
  -real &Use reals instead of integers as the default type.\\
  -redlog &Call Redlog for base prover. Assumes '-real'. \\ 
  -version &Print version number.\\
\hline
\end{tabular}
\footnotetext[5]{This is automatically carried out if we have a multiple-step extension. This is because the next step can only be carried out if the current
                          step resulted in ground clauses.}

\footnotetext{This is because CVC only provides booleans as bit-vectors of length 1. The type 'BOOLEAN' is the type of formulas. }

\noindent
\begin{tabular}{||ll||} \hline
   -freeType &Enables the use of an unspecified type 'free' \\
            & in addition to 'real' and 'int'.\\
            &Only CVC, Z3 and Yices accept free types. \\
            &Yices is default.\\ 
 -renameSubformulas &Toggles the renaming of subformulas\\
         &during clausification (true/false).\\
         &Subformula renaming  avoids exponential growth.\\
	 &Default is ``true''.\\
  -verbosity &Verbosity level (0,1,2).\\
         & From taciturn to garrulous.\\
	 &To be used in conjunction with '-prClauses'.\\
         &Default is 0\\
 -arrays   &Use settings for array.\\
           &This combines '-preprocess', '-min' and '-arith';\\
            & It also splits clauses on negative equalities.\\
  -model    & Gives a counter-model for satisfiable proof tasks.\\
            & Needs Yices or CVC (implies Yices by default).\\
  -smt      &Produce SMT-LIB output\\
            & without calling a prover.\\
  -isLocal  & Use this flag (true/false) to tell the program\\ 
            & whether all the  extensions are local or not.\\
            & This matters only if \hpilot\\
            & cannot derive a contradiction.\\
            & In that case this means that there really is none \\
            & only if the extensions are local.\\
            & Default is false.\\
  -help  &Display  list of options\\
 \hline
\end{tabular}

\section{Error handling}

\noindent In case there is a parsing error one can use
\medskip

\verb+export OCAMLRUNPARAM='p'+ (in $bash$ syntax).

\medskip
\noindent This produces a walk-through of the parsing process, which 
is of great help in localizing syntax errors.
To turn it back off use

\medskip
\verb+export OCAMLRUNPARAM=''+.

\section{Application areas}

H-PILoT has applications in \emph{mathematics}, \emph{multiple-valued
  logics}, \emph{data-} \emph{structures} and \emph{reasoning in complex
  systems}.

\smallskip
\noindent{\bf Mathematics.} An important example of local extensions
are extensions with \emph{monotone functions} over partially ordered
sets \cite{sofronie-ihlemann-ismvl-07,sofronie-ihlemann-jmvl-07}.
  We will give an example of how to use H-PILoT on problems
involving monotonicity in Section~\ref{sec-examples} below. Another
example from mathematics is an extension with \emph{free functions},
i.e., we have an empty set of extension clauses $\K$ but the proof
task $G$ contains new function symbols. Even in this simple case,
local and hierarchical reasoning is useful, because expanding the
signature might already derail a back-end prover. For instance,
consider real arithmetic. Linear arithmetic is tractable and can be
handled quite efficiently by state-of-the-art SMT-solvers.  Non-linear
arithmetic is another matter, however (cf. \cite{jsat,absolver}). Here
the options are more limited. To handle non-linear real arithmetic,
H-PILoT is integrated with the prover \emph{Redlog}
(\cite{redlog}). Since Redlog uses quantifier elimination for real
closed fields, it relies on a fixed signature. In a case like this,
H-PILoT can be employed as a preprocessor that eliminates the new
function symbols in the proof task, allowing the user to employ free
function symbols together with (non-linear) real arithmetic.
Another example from mathematics, taken from
\cite{sofronie-cade05}, is that of a \emph{Lipschitz function}.  
There it was shown
that any extension of the real numbers with a Lipschitz function is
local.

\smallskip
\noindent{\bf Multiple-valued logics.} Another important application
area of local reasoning and, by the same token, H-PILoT is reasoning
in multiple-valued logics.  These logics have more than two truth
values, in fact they allow as truth values the whole real interval of
$[0, 1]$.  The semantics are often given algebraically. For example,
the class ${\cal MV}$ of all MV-algebras is the quasi-variety
generated by the real unit interval $[0, 1]$ with connectives $\{
\vee, \wedge, \circ, \Rightarrow \}$
(cf.~\cite{sofronie-ihlemann-ismvl-07,sofronie-ihlemann-jmvl-07}).
The connectives for these algebras can be defined in terms of real
functions and relations. Hence, these connectives can be seen as the
extension functions of a definitional extension over the reals, which
is local.  One may, therefore, reduce universal validity problems over
the class of ${\cal MV}$-algebras, say, to a constraint satisfiability
problem over the unit interval $[0, 1]$
(cf. \cite{sofronie-ihlemann-ismvl-07,sofronie-ihlemann-jmvl-07}). This
allows one to use solvers for the real numbers to discharge proof
tasks over multiple-valued logics.  We will give an example of this in
Section \ref{sec-mv}.

\smallskip
\noindent{\bf Data structures.}
The ubiquitous data structures of arrays and lists satisfy locality
conditions if we confine ourselves to appropriate fragments of their
theories.  This matters in particular if we have satisfiable
problems. In order to have a full decision procedure - one that is
also able to give the correct answer for satisfiable problems  - one
has to stay inside of these fragments.  H-PILoT automates this task:
it will check whether a given problem lies inside the appropriate
fragment of the theory of arrays or pointers, respectively, and give
the answer ``satisfiable'' only if this is the case.  Otherwise,
H-PILoT will give the answer ``unknown'' and warn the user that the
problem did not fall inside the local fragment.  (For unsatisfiable
problems this never matters. If we can derive a contradiction from the
local instances alone, we can derive one from the universal extension
axioms a fortiori.)

\smallskip
\noindent{\bf Reasoning in complex systems.}
In order to be able to handle complex real life systems which mix many
theories, a stratified approach is expedient: we consider \emph{chains
  of local extensions} (cf. \cite{sofronie-jacobs:verification}).  This
feature is supported in H-PILoT. The user simply has to tag extension
functions with their respective level in the chain. A reduction is
then carried out iteratively by the program.  A full-fledged reduction
is possible provided the reduced theory clauses are ground at each
level of the extension chain.  H-PILoT has been part of a vertically
integrated tool chain, checking invariants of a transition system
modeling a European train controller system (see
Section~\ref{sec-case-study}).  The correctness of the model was shown
automatically \cite{sofronie-ihlemann-jacobs:tacas}.  The
underlying track topology was complex and dynamic, making H-PILoT's
ability to decide the pointer fragment essential. It was also a great
help in practice, due to its ability to provide (readable)
counterexamples in the cases where the problem together with the
axiomatization was satisfiable. This aided the modeler in finding gaps
in the specification.

\section{Examples}\label{sec-examples}

We illustrate the way \hpilot is implemented and can be used on two
examples. 
\subsection{Monotone functions} 
We consider as base theory $\T_0$ the theory of a partial order, 
and its extension with two monotone functions $f$ and $g$. 

That is, our
base theory $\T_0$ consists of the axioms for reflexivity, transitivity and
anti-symmetry.

\begin{enumerate-}
\item $\forall x.\; x \leq x$.
\item $\forall x,y.\; x \leq y \wedge y \leq x \rightarrow x = y$.
\item $\forall x,y,z.\; x \leq y \wedge y \leq z \rightarrow x \leq z$.
\end{enumerate-}

The extension we consider consists of the two new function symbols together with the clauses $\K$ expressing their monotonicity.

\begin{enumerate-}
\item $\forall x,y.\; x \leq y \rightarrow f(x) \leq f(y)$.
\item $\forall x,y.\; x \leq y \rightarrow g(x) \leq g(y)$.
\end{enumerate-}


\smallskip
\noindent We want to show that
\[ {{\T}_0 \cup \K} \models c_0 \leq c_1 \leq d_1 \wedge c_2 \leq d_1 \wedge d_2 \leq c_3 \wedge d_2 \leq c_4 \wedge f(d_1) \leq g(d_2)  \rightarrow f(c_0) \leq g(c_4).\]
Expressed as a satisfiability problem of the form ``${\T}_0 \cup \K \cup G \models \perp?$'', where:
$$G = c_0 \leq c_1 \leq d_1 \wedge c_2 \leq d_1 \wedge d_2 \leq c_3 \wedge d_2 \leq c_4 \wedge f(d_1) \leq g(d_2) \wedge \neg f(c_0) \leq g(c_4).$$

\noindent As an input file for \hpilot this looks as follows (we use ``R'' as order relation because $\leq$ is reserved).

\begin{MyFVerbatim}
% Two monotone functions over a poset.
% Status: unsatisfiable

Base_functions:={}
Extension_functions:={(f, 1), (g, 1)}
Relations:={(R, 2)}

% R is partial order
Base := (FORALL x).     R[x, x];
        (FORALL x,y,z). R[x, y], R[y, z] --> R[x, z];
        (FORALL x,y).   R[x, y], R[y, x] --> x = y;

Clauses := (FORALL x,y). R[x, y] --> R[f(x), f(y)];
           (FORALL x,y). R[x, y] --> R[g(x), g(y)];

Query := R[c0, c1];
         R[c1, d1];
         R[c2, d1];
         R[d2, c3];
         R[d2, c4];
         R[f(d1), g(d2)];
         NOT(R[f(c0), g(c4)]);
\end{MyFVerbatim}

\noindent In this case we have no function symbols  in the base theory and two functions symbols $f$ and $g$ of arity 1 in the extension clauses.
This is expressed by:

\begin{MyVerbatim}
Base_functions:={}
Extension_functions:={(f, 1), (g, 1)}
\end{MyVerbatim}

We have only one relation in our (base) clauses, namely 'R', with arity 2.
This we express by: 

\medskip
\verb+Relations:={(R, 2)}+.
\medskip

For technical reasons, relations require square brackets for their arguments in \hpilot as seen above.
The symbols \verb'<=', \verb'<', \verb'>=' and \verb'>' are reserved for arithmetic over the integers or over the reals.
They may be written infix and there are provers (e.g., Yices, CVC) that ``understand'' arithmetic and orderings.
We wouldn't have needed to axiomatize '$\leq$' at all. 

However, the above problem is more general. It concerns every partial
order not only orderings of numbers.  By default, \hpilot calls
SPASS. SPASS has no in-built understanding of orderings and, thus,
\verb'<=' would be just an arbitrary symbol.  For clarity we used the
letter 'R'.

As for the syntax of clauses, one should note that the syntax of
\hpilot requires that each clause must end with a semicolon, be in
prenex normal form and all names meant to be (universal) variables
must be explicitly quantified.

\medskip
\noindent
\begin{centering}
\fbox{{\it Every name which is not explicitly quantified will be considered a constant!}}
\end{centering}

\bigskip

\noindent
As we can see in our proof task.

\begin{MyVerbatim}
Proof Task := R[c0, c1];
         R[c1, d1];
         R[c2, d1];
         R[d2, c3];
         R[d2, c4];
         R[f(d1), g(d2)];
         NOT(R[f(c0), g(c4)]);
\end{MyVerbatim}

Note further that because the background theory, extension theory and
the proof task must all be clauses\footnote{In fact, the extension clauses
  might be more general as we will see later.}, we need to break up
the conjunction in our original proof task into a set of unit clauses.  A
non-unit clause may be written as above in a ``sorted'' manner
$\varphi_1, ...,\varphi_n \rightarrow \psi_1, ..., \psi_k$ for
$\varphi_1 \wedge ... \wedge \varphi_n \rightarrow \psi_1 \vee ...\vee
\psi_k$, i.e., as an implication with the negated atoms of the clause
in the antecedent and the (positive) atoms in the consequent (the
operator \verb+-->+ is reserved for sorted clauses) or as an
arbitrary disjunctions of literals.

The name of the input files for \hpilot can be freely chosen, although it is customary to have them have the suffix ``.loc''.
Suppose we have put the above problem in a file named \verb+mono_for_poset.loc+, then we can run \hpilot by calling

\begin{MyVerbatim}
hpilot.opt mono_for_poset.loc
\end{MyVerbatim}

\hpilot will parse the input file, carry out the reduction and then will hand over the reduced problem to SPASS
(using the same name but with the suffix ``.dfg'').
SPASS terminates quickly with the result that a proof exists

\begin{MyFVerbatim}[frame=lines]
SPASS beiseite: Proof found.
Problem: mono_for_poset.dfg
SPASS derived 35 clauses, backtracked 0 clauses and kept 41 clauses.
SPASS allocated 496 KBytes.
SPASS spent     0:00:02.32 on the problem.
                0:00:00.00 for the input.
                0:00:00.00 for the FLOTTER CNF translation.
                0:00:00.00 for inferences.
                0:00:00.00 for the backtracking.
                0:00:00.10 for the reduction.
\end{MyFVerbatim}

\noindent meaning that the set of clauses is inconsistent, as we wanted to show.

\medskip
\noindent One can see the full reduction process by using the option \verb+-prClauses+.

\subsection{Arrays}
For a more complicated example, let us consider an algorithm for
inserting an element $x$ into a sorted array $a$ with the bounds $l$
and $u$.  We want to check that the algorithm is correct, i.e., that
the updated array $a'$ remains sorted. This could be an invariant
being checked in a verification task. 

There are three different cases. 
\begin{itemize}
\item $x$ could be smaller than any element in $a$
(equivalently, $x < a[l]$), 
\item $x$ could be greater than any element of
$a$ ($x > a[u]$) or, 
\item there is a position $p$ ($l < p \leq u$)
such that $a[p-1] < x$ and $x \leq a[p]$.  
\end{itemize}
In the first two cases we
put $x$ at the first respectively last position of the array.  In the
third case, we insert $x$ at position $p$ and shift the other elements
to the right, i.e., $a'[i+1] = a[i]$ for $i > p$.  We have to take
care to cover also the cases where the array contains only 1 or 2
elements.  As input it will look as follows.

\begin{MyFVerbatim}
Clauses := 
  % case 1
  (FORALL i). i = l, x <= a(i) --> a'(i) = x;
  (FORALL i). x <= a(l), l < i, i <= u + _1 --> a'(i) = a(i - _1);

  % case 2
  (FORALL i). i = u, a(i) <= x --> x <= a(l), a'(i + _1) = x;
  (FORALL i). a(u) <= x, l - _1 <= i, i < u 
                       --> x <= a(l), a'(i + _1) = a(i + _1);

  % case 3
  (FORALL i). x < a(u), l <= i, i < u, a(i) < x, x <= a(i + _1) 
                 --> a'(i + _1) = x;
  (FORALL i). a(l) < x, x < a(u), l <= i, i < u, x <= a(i), 
                 x <= a(i + _1) --> a'(i + _1) = a(i);
  (FORALL i). a(l) < x, x < a(u), i = u + _1 --> a'(i) = a(i - _1);  
  (FORALL i). a(l) < x, x < a(u), l - _1 <= i, i < u, a(i + _1) < x 
                 --> a'(i + _1) = a(i + _1);
          
  (FORALL i,j). l <= i, i <= j, j <= u --> a(i) <= a(j);
\end{MyFVerbatim}

\noindent with the last clause saying that $a$ was sorted at the beginning.

There are several things to note. Most importantly, we now have a
two-step extension. First, an array can be simply seen as a partial
function. This gives us the first extension $\T_0 \subseteq \T_1$. $\T_0$
here is the theory of indices (integers, say) which we extend by the
function $a$ and the axiom for monotonicity of $a$. Now we update $a$,
giving us a second extension $\T_2 \supseteq \T_1$ where our extension
clauses $\K_2$ are given by the three cases above.

We need to make sure that the last extension is also
local. This is easy to establish, because $\K_2$ is a
\emph{definitional extension} or case distinction
(cf.~\cite{sofronie-ihlemann-ismvl-07,sofronie-ihlemann-jmvl-07}). 
A definitional extension is one
where extension functions $f$ only appear in the form $\varphi_i(\x)
\rightarrow f(\x) = t_i(\x)$ with $t_i$ being a base theory term and
the $\varphi_i$ are mutually exclusive base theory clauses.  This is
the reason that we have written $\forall i. i = l, x \leq a(i)
\rightarrow a'(i) = x$ instead of the shorter $x \leq a(l) \rightarrow
a'(l) = x$: to ensure that the antecedents of the clauses are all
mutually exclusive.  Now we know that we are dealing with a
definitional and therefore local extension.  (Remember that when 
assessing whether $\T_2 \supseteq \T_1$ is a local extension, 
$\T_1$ is the base theory; that $\T_1$ is itself an extension is 
not important at the moment.)

We need to tell the program that we are dealing with a chain of
extensions instead of a single one. We do this by declaring to 
which level of the chain an extension function
belongs: $(f, arity, level)$.

In our example that would be
\begin{MyVerbatim}
Extension_functions:={(a', 1, 2), (a, 1, 1)}
\end{MyVerbatim}

The program will now automatically determine the level of each
extension clause. In our example, an extension clause will have level
2 if and only if $a'$ occurs in it and level 1 otherwise (level 0
refers to a clause in the base theory).

Also recall from the explanations in Section~\ref{parameters} 
that numerals (names for integers) must be preceded by an
underscore, and that $+$ and $-$ may be written infix
for readability; ($=, +, -, *, /$) are the only functions for which
this is allowed\footnote{When using SPASS, they may also be written
  infix but nevertheless they are just free functions for SPASS.}.

 Our declarations, therefore should look like this.

\begin{MyVerbatim}
Base_functions:={(+,2), (-, 2)}
Extension_functions:={(a', 1, 2), (a, 1, 1)}
Relations:={(<=, 2), (<, 2)}
\end{MyVerbatim}

\noindent All that is left to do now is add the proof task -- the negation of
$$\forall i,j.\; (l \leq i \leq j \leq u + 1\rightarrow a'(i) \leq a'(j))$$ 
namely: 
$$ 1 \leq m \wedge m \leq n \wedge n \leq u + 1 \wedge \neg (a'(m) \leq a'(m))$$
to the file and hand it over to \hpilot.
The file looks like this.

\begin{MyFVerbatim}%[frame=lines]
% ai.loc
% Arrays for definitional extensions; 

Base_functions:={(+,2), (-, 2)}
Extension_functions:={(a', 1, 2), (a, 1, 1)}
Relations:={(<=, 2), (<, 2)}

% K
Clauses := 
  % case 1
  (FORALL i). i = l, x <= a(i) --> a'(i) = x;
  (FORALL i). x <= a(l), l < i, i <= u + _1 --> a'(i) = a(i - _1);

  % case 2
  (FORALL i). i = u, a(i) <= x --> x <= a(l), a'(i + _1) = x;
  (FORALL i). a(u) <= x, l - _1 <= i, i < u 
                       --> x <= a(l), a'(i + _1) = a(i + _1);

  % case 3
  (FORALL i). x < a(u), l <= i, i < u, a(i) < x, x <= a(i + _1) 
                 --> a'(i + _1) = x;
  (FORALL i). a(l) < x, x < a(u), l <= i, i < u, x <= a(i), 
                 x <= a(i + _1) --> a'(i + _1) = a(i);
  (FORALL i). a(l) < x, x < a(u), i = u + _1 --> a'(i) = a(i - _1);  
  (FORALL i). a(l) < x, x < a(u), l - _1 <= i, i < u, a(i + _1) < x 
                 --> a'(i + _1) = a(i + _1);
          
  (FORALL i,j). l <= i, i <= j, j <= u --> a(i) <= a(j);

Query := l <= m;
         m <= n;
         n <= u + _1;
         NOT( a'(m) <= a'(n) );
\end{MyFVerbatim}

We do not need to declare a base theory here because we will be using Yices and Yices already ``knows'' integer arithmetic. We
call \hpilot thus: 

\begin{MyVerbatim}
hpilot.opt -yices -preprocess ai.loc
\end{MyVerbatim}

 \hpilot will produce a reduction, put it in a file called $ai.ys$ and pass it over to Yices
which will say \verb+unsat+ or \verb+sat+. A note on the flag \verb+-preprocess+: Establishing that some extension
is local presupposes that the extension clauses in $\K$ are \emph{flat} and \emph{linear}. 
Flatness means that the clauses contain no nesting
of extension functions.  
Linearity means that:  
\begin{itemize}
\item no variable occurrs twice in any extension term and 
\item if any variable occurs in two extension terms, the terms are the same. 
\end{itemize}
In this example, we have non-flat clauses such as 
\begin{MyVerbatim}
(FORALL i). i = u, a(i) <= x --> x <= a(l), a'(i + _1) = x;
\end{MyVerbatim}
We rectify matters by a \emph{flattening} operation - rewriting the above clause to
\begin{MyVerbatim}
(FORALL i,j). j = i + _1, i = u, a(i) <= x --> 
                                      x <= a(l), a'(j) = x;
\end{MyVerbatim}
This will not affect consistency of any proof task w.r.t. $\K$. 
However, it does not improve readability.
Therefore the program will perform flattening/linearization only 
if the option \verb+-preprocess+ is chosen.

\section{Example: Specifying the type information}
\label{types}
\subsection[Global Constraints]{Global constraints\footnote{ This feature is not supported for Z3.}}
\label{sec-mv}

Sometimes we want to restrict the domain of the problem, e.g., we want to consider natural numbers instead of integers
or we are interested in a real interval $[a, b]$ only. Yices and CVC support the definition of subtypes. When using
one of these it is possible to state a global constraint on the domain of the models in the preamble as follows:

\begin{MyVerbatim}
Interval := 0 <= x <= 1;
\end{MyVerbatim}

\noindent This will restrict the domain of the models of the theory to
the unit interval $[0, 1]$. It is equivalent to adding the antecedent
$0 \leq x \wedge x \leq 1$, for every variable $x$, to each formula in
the clauses and the proof task.

The bounds of the interval
can also be exclusive or mixed as in

\begin{MyVerbatim}
Interval := 0 < x <= 1;
\end{MyVerbatim}

\noindent
or one-sided as in

\begin{MyVerbatim}
Interval := 2 <= x;
\end{MyVerbatim}

Consider the following example, taken from
\cite{sofronie-ihlemann-ismvl-07,sofronie-ihlemann-jmvl-07}, concerning multiple-valued
logic.  The class ${\cal MV}$ of all MV-algebras is the quasi-variety
generated by the real unit interval $[0, 1]$ with the \L ukasiewicz
connectives $\{ \vee, \wedge, \circ, \Rightarrow \}$, i.e., the
algebra $[0, 1]_{{L}} = ([0, 1], \vee, \wedge, \circ, \Rightarrow)$.
The {\L}ukasiewicz connectives can be defined in terms of the
real functions '$+$','$-$' and the relation '$\leq$', giving us a local
extension over the real unit interval.

Therefore, the following are equivalent:
\begin{itemize}
\item[(1)]  ${\cal MV} \models \forall {\overline x} \bigwedge_{i = 1}^n s_i({\overline x})  = t_i({\overline x}) \rightarrow s({\overline x}) = t({\overline x})$

\item[(2)] $[0, 1]_{{L}}  \models \forall {\overline x} \bigwedge_{i = 1}^n s_i({\overline x}) = t_i({\overline x}) \rightarrow s({\overline x}) = t({\overline x})$

\item[(3)] $\T_0 \cup {\sf Def}_{{L}} \wedge \bigwedge_{i = 1}^n s_i({\overline c}) = t_i({\overline c}) \wedge s({\overline c}) \neq t({\overline c}) \models \perp$, 
\end{itemize}
where $\T_0$ consists of the real unit interval $[0, 1]$  with the 
operations $+, -$ and predicate symbol $\leq$.

\medskip

 For instance, we might want to establish whether linearity  
$(x \Rightarrow y) \vee (y \Rightarrow x) = 1$ follows from the axioms.
As an input file for \hpilot it looks like this.

\begin{MyFVerbatim}%[frame=lines]
% file mv1.loc
% Example for MV-algebras
% The Lukasiewicz connectives can be defined 
% in terms of the (real) connectives +,-,<=

Base_functions:={(+, 2), (-, 2)}
Extension_functions:={(V, 1), (M, 1), (o, 1), (r, 1)}
Relations:={(<=, 2), (<, 2), (>, 2), (>=, 2)}

Interval := 0 <= x <= 1;

% K
Clauses := % definition of \/
           (FORALL x,y). x <= y --> V(x, y) = y;
           (FORALL x,y). x > y  --> V(x, y) = x;

           % definition of /\
           (FORALL x,y). x <= y --> M(x, y) = x;
           (FORALL x,y). x > y  --> M(x, y) = y;

           % definition of o
           (FORALL x,y). x + y <  _1 --> o(x, y) = _0;
           (FORALL x,y). x + y >= _1 --> o(x, y) = (x + y) - _1;

           % definition of =>
           (FORALL x,y). x <= y --> r(x, y) = _1; 
           (FORALL x,y). x > y  --> r(x, y) = (_1 - x) + y;

Query := % linearity: x => y . \/ . y => x = 1
         NOT(V(r(a, b), r(b, a)) = _1);
\end{MyFVerbatim}

\subsection{Using standard types}

In default mode using SPASS, \hpilot hands over a set of general
first-order formulas without types.  However, \hpilot also provides
support for the standard types \verb+int, real, bool+ and for free
types.  When using CVC or Yices the default type is \verb+int+, for
Redlog it is \verb#real#. The default type does not need to be
specified in the input file.  One can also use the \verb+-real+ flag
to set the default type to $\mathrm{real}$ for Yices and CVC.

Free types are specified as \verb+free#+$i,\; i=1,2,\dots$ or simply as
\verb+free+ if there is only one free type.  When using free types the
flag \verb+-freeType+ must be set.  Only Yices and CVC are able to
handle free types (Yices is default when the flag is set).  When using
mixed type in one input file, the types of the functions and the
constants need to be declared.
If the domain of a function is the same as the range it is enough to specify the domain as in

\medskip
$(foo, arity, level, domainType)$
\medskip

\noindent if they differ say

\medskip
$(foo, arity, level, domainType, rangeType).$
\medskip

\noindent Constants are simply declared as
\medskip

$(name, type).$
\medskip

\noindent The following example is taken from \cite{sofronie:local-reasoning}.

\begin{MyFVerbatim}
% Pointers
% status unsatisfiable
Base_functions:={(+,2), (-, 2)}
Extension_functions:={(next, 1, 1, free#1), (prev, 1, 1, free#1), 
                         (priority, 1, 1, free#1, real), 
                            (state, 1, 1, free#1, free#2)}
Relations:={(>=, 2)}
Constants:={(null, free#1), (eps, real), (a, free#1), (b, free#1), 
               (RUN, free#2), (WAIT, free#2), (IDLE, free#2)}

% K
Clauses := 
  (FORALL x). OR(state(x) = RUN, state(x) = WAIT, state(x) = IDLE);
  % prev and next are inverse
  (FORALL p). OR(p = null, prev(next(p)) = null, prev(next(p)) = p);
  (FORALL p). --> p = null, next(prev(p)) = null, next(prev(p)) = p;
  (FORALL p, q). next(q) = next(p) --> p = null, q = null, p = q; 
  (FORALL p, q). prev(q) = prev(p) --> p = null, q = null, p = q; 
  (FORALL p). --> p = null, next(p) = null, state(p) = IDLE, 
                    state(next(p)) = IDLE, state(p) = state(next(p));
  (FORALL p). OR(p = null, next(p) = null, NOT(state(p) = RUN), 
                      priority(p) >= priority(next(p)));


Query := NOT(eps = _5);
         NOT(eps = _6);
         priority(a) = _5;
         priority(b) = _6;
         a = prev(b);
         state(a) = RUN;
         NOT(next(a) = null);
         NOT(a = null);
         NOT(b = null);
\end{MyFVerbatim}

\section{Example: Handling data structures}
\subsection{Arrays}\label{sec-arrays}

We consider the local fragment of the theory of arrays in more detail and  show how it can be dealt with by \hpilot$\!\!$.
For the \emph{array property fragment} (\cite{bradley-manna}) the following syntactical
restrictions are imposed. Let $A$ be a set of function symbols used for 
denoting arrays.

\begin{enumerate}
\item An \emph{index guard} is a positive Boolean combination of atoms of the form $t \leq u$ or $t = u$ where $t$ and $u$
      are either a variable or a ground term of linear arithmetic.
\item A {\em value restriction} is a formula $\varphi_V({\overline c}, {\overline x})$ containing constants among those in ${\overline c} = c_1, \dots, c_k$ 
and free variables among those in 
${\overline x} = x_1, \dots, x_n$, with the property that:
\begin{itemize}
\item[(1)]  all occurrences of the variables are shielded by function symbols in $A$; 
\item[(2)] no nested array reads are allowed
\end{itemize}
i.e., the free variables $x_i$ occur in $\varphi_V$ only 
in direct array reads $a[x_i]$. 
\item A universal formula of the form $(\forall \x)(\varphi_I(\x) \rightarrow \varphi_V(\x))$ is an \emph{array property} if 
it is flat, $\varphi_I$ is an index guard and $\phi_V$ a value restriction.
\end{enumerate}

In this section we only consider extensions by clauses of the above form. 
Our base theory is the disjoint, many-sorted combination of linear
integer arithmetic (Presburger) with a theory of elements. The extension 
functions are in this case the function symbols in $A$, used for denoting 
arrays.
In order to be able to handle this fragment we have to use a particular type 
of locality, namely minimal locality. 
To use this feature we call \hpilot with parameter \verb+-arrays+: 

\medskip
\verb+hpilot.opt -arrays k.loc+
\medskip

\medskip
Consider the example of inserting a new element into a sorted array $a$.
Arrays are modeled as free functions and array updates are dealt with by introducing new array names.
In this fashion, let $d$ be identical to $a$ except for position $k$ at 
which it takes value $w$ 
and let $e$ be identical to $d$ except possibly at position $l$ where we have 
written $x$ and similarly for $c$, $b$ and $a$.
The set $\K$ of extension clauses we consider is: 
\begin{equation*}
\left.\begin{aligned}
&(\forall i,j)(0 \leq i \leq j \leq n-1 \rightarrow c[i] \leq c[j] ) &(1)&\\
&(\forall i,j)(0 \leq i \leq j \leq n-1 \rightarrow e[i] \leq e[j] ) &(2)&\\
&(\forall i)(i \neq l \rightarrow b[i] = c[i] ) &(3)&\\
&(\forall i)(i \neq k \rightarrow a[i] = b[i] ) &(4)&\\
&(\forall i)(i \neq l \rightarrow d[i] = e[i] ) &(5)&\\
&(\forall i)(i \neq k \rightarrow a[i] = d[i] ). &(6)&\\
\end{aligned}
\right\}
\quad \K
\end{equation*}

\noindent
Our proof task (with additional constraints) is
\begin{equation*}
\left.\begin{aligned}
&w < x < y < z\\
&0 < k < l < n\\
&k + 3 < l\\
&c[l] = x\\
&b[k] = w\\
&e[l] = z\\
&d[k] = y.
\end{aligned}
\right\}
\quad G
\end{equation*}

\noindent The input file looks as follows.
(The operators are written prefix here which requires the names \verb+plus+ and \verb+minus+,
because \verb_+_ and \verb_-_ are reserved for infix.)


\begin{MyFVerbatim}
% Arrays for minimal locality
Base_functions:={(plus,2), (minus, 2)}
Extension_functions:={(a, 1), (b, 1), (c, 1), (d, 1), (e, 1)}
Relations:={(<=, 2)}

% K
Clauses := (FORALL i,j). _0 <= i, i <= j, 
                             j <= minus(n, _1) --> c(i) <= c(j);
           (FORALL i,j). _0 <= i, i <= j, 
                             j <= minus(n, _1) --> e(i) <= e(j);
           (FORALL i).  --> i=l, b(i) =  c(i);
           (FORALL i).  --> i=k, a(i) =  b(i);
           (FORALL i).  --> i=l, d(i) =  e(i);
           (FORALL i).  --> i=k, a(i) =  d(i);


Query := plus(w, _1)  <= x;
         plus(x, _1)  <= y;
         plus(y, _1)  <= z;
         plus(_0, _1) <= k;
         plus(k, _1)  <= l;
         plus(l, _1)  <= n;
         plus(k, _3)  <= l;
         c(l) = x;
         b(k) = w;
         e(l) = z;
         d(k) = y;
\end{MyFVerbatim}

$\K$ does not yet fulfil the syntactic requirements (index guards must be positive!).
We rewrite $\K$ as follows: We change an expression $i \neq l$ where $i$ is the (universally quantified) variable to $i \leq l-1 \vee l+1 \leq i$.
We rewrite it like this because the universally quantified variable 
$i$ must appear unshielded in the index guard. This gives us the following set
of clauses.
\begin{equation*}
\left.\begin{aligned}
&(\forall i,j)(0 \leq i \leq j \leq n-1 \rightarrow c[i] \leq c[j] ) &(1)&\\
&(\forall i,j)(0 \leq i \leq j \leq n-1 \rightarrow e[i] \leq e[j] ) &(2)&\\
&(\forall i)(i \leq l-1 \rightarrow b[i] =  c[i] ) &(3)&\\
&(\forall i)(l+1 \leq i \rightarrow b[i] =  c[i] ) &(4)&\\
&(\forall i)(i \leq k-1 \rightarrow a[i] =  b[i] ) &(5)&\\
&(\forall i)(k+1 \leq i \rightarrow a[i] =  b[i] ) &(6)&\\
&(\forall i)(i \leq l-1 \rightarrow d[i] =  e[i] ) &(7)&\\
&(\forall i)(l+1 \leq i \rightarrow d[i] =  e[i] ) &(8)&\\
&(\forall i)(i \leq k-1 \rightarrow a[i] =  d[i] ) &(9)&\\
&(\forall i)(k+1 \leq i \rightarrow a[i] =  d[i] ). &(10)&\\
\end{aligned}
\right\}
\quad \K'
\end{equation*}
\hpilot performs this and the following rewrite steps automatically
to spare the user this tedious labor.
Also, $\K$ is not linear, this must also be taken care of. \hpilot carries 
out all necessary rewrite steps
for the user, who can simply input the above file to the system.

\medskip
Instead of using free functions to specify array updates,
\hpilot allows the user to {\em model array updates directly} by using
a ``write'' function -- 
for example, ${\rm write}(a, i, x)$ denotes a new array
which is identical to $a$ except (possibly) at position $i$ where 
the value of the new array is set to $x$.
In this way we can specify our problem above as: 

\begin{MyFVerbatim}
% Arrays for minimal locality with 'write'function.
Base_functions:={(+, 2), (-, 2)}
Extension_functions:={(a, 1)}
Relations:={(<=, 2)}

% K
Clauses := 
 (FORALL i,j). _0 <= i, i <= j, j <= n - _1 --> 
    write(write(a,k,w), l, x)(i) <= write(write(a,k,w), l, x)(j);

 (FORALL i,j). _0 <= i, i <= j, j <= n - _1 --> 
    write(write(a,k,y), l, z)(i) <= write(write(a,k,y), l, z)(j);

Query := w + _1  <= x;
         x + _1  <= y;
         y + _1  <= z;
         _0 + _1 <= k;
         k + _1  <= l;
         l + _1  <= n;
         k + _3  <= l;
\end{MyFVerbatim}

As above, \hpilot will also automatically split on disequations in the antecedent.
Note also that since we assume that indices of arrays are integers, it makes no difference whether we write
\verb@w + _1@ or \verb+plus(w, _1)+ in the input file. Linear integer arithmetic will be used (Yices is default).

\subsection{Pointers}
\label{sec-pointers}

The local fragment of the theory of pointers (cf. \cite{necula-mcpeak,sofronie-ihlemann-jacobs:tacas,faber-ihlemann-jacobs-sofronie-ifm10}) is also implemented in \hpilot$\!\!$.
We consider pointer problems over a two-sorted language, containing one sort 
\verb'pointer' and another
scalar sort. The scalar sort can be concrete e.g.\ \verb'real', or is 
kept abstract in which case it is written as \verb'scalar'.
There are  two function types involving pointers, namely $\verb+pointer+ \rightarrow \verb+pointer+$ and $\verb+pointer+ \rightarrow scalar$,
where $scalar$ is either a concrete scalar sort (e.g.\ \verb+real+) or the 
abstract sort ``scalar''.\footnote{In 
\cite{faber-ihlemann-jacobs-sofronie-ifm10} we use an extension of 
\hpilot which allows several pointer sorts, as well as functions of sort 
$p_1, \dots p_n \rightarrow p$ (where $p_i, p$ are pointer sorts) and 
$p_1,\dots, p_m \rightarrow scalar$. This aspect is discussed at the end of 
this section.} 
%
The axioms we consider are all of the form 
\begin{eqnarray*}
\forall \bar{p}. ~~ {\cal E} \vee {\cal C}
\end{eqnarray*}

\noindent
where $\bar{p}$ is a set of pointer variables containing all the pointer variables 
occurring in ${\cal E} \vee {\cal C}$, ${\cal E}$ contains disjunctions of pointer equalities and
${\cal C}$ is a disjunction of scalar constraints (i.e.,\ literals of scalar type). 
${\cal E} \vee {\cal C}$ may also contain free variables of scalar type or, equivalently, free scalar constants.

We require that pointer terms appearing below a function should not be ${\sf null}$ in order to rule out null pointer errors.
That is, for all terms $f_1(f_2(\dots f_n(p)))$, $\, i= 1,..,n$, occurring in 
the  axiom, the axiom also contains 
the disjunction $p = {\sf null} \vee f_n(p) = {\sf null} \vee \dots \vee 
f_2(\dots f_n(p)) = {\sf null}$.

Pointer/scalar formulas complying with this restriction are called 
\emph{nullable}.
The locality result in \cite{sofronie-ihlemann-jacobs:tacas}
 allows the integration of pointer reasoning with the above features
into H-PILoT.
We now present an example given in \cite{necula-mcpeak}, which 
looks like this as input for
\hpilot$\!\!$. (We have added an appropriate proof task.)


\begin{MyFVerbatim}
Base_functions:={(+,2), (-, 2)}
Extension_functions:={(next, 1, 1, pointer), 
                      (prev, 1, 1, pointer), 
                      (priority, 1, 1, pointer, real)}
Relations:={(>=, 2)}
Constants:={(a, pointer), (b, pointer)}

Clauses := 
     (FORALL p). prev(next(p)) = p;
     (FORALL p). --> next(prev(p)) = p;
     (FORALL p). --> q = null, priority(p) >= priority(next(p));

Query := priority(a) = _5;
         priority(b) = _6;
         a = prev(b);
         NOT(a = null);
         NOT(b = null);
\end{MyFVerbatim}

\medskip

\hpilot can be called without any parameters because the keyword
\verb+pointer+ is present.  This will trigger H-PILoT's pointer mode
so that it will add all the nullable terms to the axioms and use the
specific (stable) locality required.

\medskip

Because the scalar type is concrete here (\verb+real+), \hpilot will
use Yices as the back-end prover (its default for arithmetic). If we
want to leave the scalar type abstract we could write something like

\begin{MyFVerbatim}
%psiPointers.scalar.loc
Base_functions:={}
Extension_functions:={(next, 1, 1, pointer), 
                      (prev, 1, 1, pointer), 
                      (priority, 1, 1, pointer, scalar)}
Relations:={}
Constants:={(a, pointer), (b, pointer), (c5, scalar), (c6, scalar)}

Clauses := (FORALL p). prev(next(p)) = p;
           (FORALL p). next(prev(p)) = p;
           (FORALL p). NOT(priority(p) = priority(next(p)));


Query := priority(a) = c5;
         priority(b) = c6;
         a = prev(b);
         c5 = c6;
         NOT(a = null);
         NOT(b = null);
\end{MyFVerbatim}

\noindent We again can simply type 

\medskip
\verb+hpilot.opt -preprocess psiPointers.scalar.loc+ 
\medskip

\noindent without any parameters. \hpilot will 
recognize this as a pointer problem and use Yices as default, this time because of the free type \verb+scalar+.
There can also be more than one pointer type and pointer extensions can be fused with other types of extensions
in a hierarchy.
However, due to the different types of locality that need to be employed, the user must specify which levels
are pointer extensions. He does this by using the keyword \verb+Stable+.

For example, the header of a more complicated verification task which mixes different pointer types might look like this.

\begin{MyFVerbatim}
Base_functions:={(-, 2), (+, 2)}
Extension_functions:=
   { % level 4
     (next',1,4, pointer#2,pointer#2), (pos',1,4,pointer#2,real)
     % level 3
     (next,1,3,pointer#2,pointer#2), (pos,1,3,pointer#2,real),
     (spd,1,3,pointer#2,real), (segm,1,3,pointer#2,pointer#1),
     % level 2
     (bd,1,2,real,real),
     % level 1
     (lmax,1,1,pointer#1,real), (length,1,1,pointer#1,real),
     (nexts,1,1,pointer#1,pointer#1), (alloc,1,1,pointer#1,int)}

Relations :={(<=, 2), (>=, 2), (>, 2), (<, 2)}

Constants:= {(t3,pointer#2), (t2,pointer#2), (t1,pointer#2),
              (d,real), (State0,int), (s,pointer#1), (State1,int)}

Stable := 1, 3;
\end{MyFVerbatim}

Note that the type \verb+pointer#2+ must be declared with a higher level 
than  \verb+pointer#1+ because
 \verb+pointer#2+ refers to  \verb+pointer#1+ but not vice versa.

\section{Example: Using the built-in CNF translator}
\label{sec-clausification}

\hpilot also provides a clausifier for ease of use. First-order
formulas can be given as input and \hpilot translates them into
clausal normal form (CNF). The CNF-translator does not provide the
full functionality of FLOTTER. It uses structural formula renaming
(\cite{plaisted-greenbaum}) and standard Skolemization, not the more
exotic variants thereof (cf. \cite{nonnengart:cnf}). Nevertheless, it
is powerful enough for most applications.
As a simple example consider the following.

\begin{MyFVerbatim}
% cnf.fol
Base_functions:={(delta, 2), (abs, 1), (-, 2)}
Extension_functions:={(f, 1)}
Relations:={}


Formulas := 
    (FORALL eps, a, x). (_0 < eps --> 
             AND( _0 < delta(eps, a), 
                     (abs(x - a) < delta(eps, a) 
                              --> abs(f(x) - f(a)) < eps)));

Query := 
\end{MyFVerbatim}

\noindent \hpilot translates  \verb+Formulas+ to clause normal form.
To see the output, we use

\begin{MyVerbatim}
     hpilot.opt -preprocess -prClauses cnf.fol
\end{MyVerbatim}

\noindent We obtain the following output file: 

\begin{MyFVerbatim}[frame=lines]
!- Adding formula: 
   (FORALL eps, a, x). 
        (_0 < eps --> 
           AND( _0 < delta(eps, a), (abs(-(x, a)) < delta(eps, a) 
             --> abs(-(f(x), f(a))) < eps)))
!- add_formulas
!- We have 1 levels.
!- done
!- Our base theory is: 
!- empty.
!- Clausifying formulas...
!- (FORALL z_1, z_3). OR( _0 < delta(z_1, z_3), NOT(_0 < z_1))
!- (FORALL z_1, z_2, z_3).
                 OR( NOT(abs(-(z_2, z_3)) < delta(z_1, z_3)), 
                     abs(-(f(z_2), f(z_3))) < z_1, NOT(_0 < z_1))
!- Yielding 2 new clauses: 
!- [z_1, z_2, z_3] abs(-(z_2, z_3)) < delta(z_1, z_3), _0 < z_1 
                       ---> abs(-(f(z_2), f(z_3))) < z_1      
!- [z_1, z_3] _0 < z_1 ---> _0 < delta(z_1, z_3)      
!- After rewriting we have as clauses K: 
!- [z_1, z_2, z_3] abs(-(z_2, z_3)) < delta(z_1, z_3), _0 < z_1 
                        ---> abs(-(f(z_2), f(z_3))) < z_1
!- [z_1, z_3] _0 < z_1 ---> _0 < delta(z_1, z_3) 
\end{MyFVerbatim}

\noindent
telling us that the above formula resulted in two new clauses 
(in addition to those given outright under \verb+Clauses+), viz.

\smallskip
~~~~~~~$\forall z_1, z_3.\;\; 0 < delta(z_1, z_3) \vee \neg(0 < z_1)$

\smallskip
and

\smallskip
$\forall z_1, z_2, z_3.\;\; \neg(abs(z_2 - z_3) < delta(z_1, z_3)) \vee abs(f(z_2) - f(z_3)) < z_1 \vee \neg(0 < z_1).$

\smallskip
\noindent In this case no ground clause resulted and \hpilot stops.

\section{Extended locality}\label{sec-extended-locality}
For some applications we would like to allow more complicated
extension clauses, say we want them to be inductive ($\forall\exists$)
instead of universal.  \hpilot is also able to handle extensions with
augmented clauses, i.e., formulas of the form $\forall x. \Phi(x) \vee
C(x)$, where $\Phi$ is an arbitrary formula \emph{which does not
  contain extension functions} and $C$ is a clause which does 
(cf.~\cite{sofronie-ihlemann-jacobs:tacas}).  Consider the following example
taken from \cite{sofronie-ihlemann-jacobs:tacas}.

Suppose there is a parametric number $m$ of processes. The priorities 
associated with the processes (non-negative real numbers) are stored in an 
array $p$. The states of the process -- enabled (1) or disabled (0) -- are 
stored in an array $a$.  At each
step only the process with maximal priority is enabled, its priority is set 
to $x$ and the priorities of the waiting processes are increased by $y$.
This can be expressed by the following set of axioms which we denote 
as ${\sf Update}(p, p', a, a')$.

\medskip

{\small 
$\begin{array}{l}
\forall i (1 \leq i \leq m \wedge ~~(\forall j (1 \leq j \leq m \wedge j {\neq} i \rightarrow p(i) {>} p(j))) \rightarrow a'(i) = 1)\quad \\
\forall i (1 \leq i \leq m \wedge ~~(\forall j (1 \leq j \leq m \wedge j {\neq} i \rightarrow p(i) {>} p(j))) \rightarrow p'(i) {=} x)  \\ 
 \forall i (1 \leq i \leq m \wedge  \neg (\forall j (1 \leq j \leq m \wedge j {\neq} i \rightarrow p(i) {>} p(j))) \rightarrow a'(i) {=} 0) \\
\forall i (1 \leq i \leq m \wedge  \neg (\forall j (1 \leq j \leq m \wedge j {\neq} i \rightarrow p(i) {>} p(j))) \rightarrow p'(i) {=} p(i) {+} y),
\end{array}$  
}

\medskip 

\noindent
where $x$ and $y$ are parameters.
We may need to check whether, given that at the beginning the priority 
list is injective, i.e.,\ formula ${\sf (Inj)}(p)$ holds: 

\medskip 

${\sf (Inj)(p)} ~~~ \forall i, j (1 \leq i \leq m \wedge 1 \leq j \leq m \wedge  i \neq j \rightarrow p(i) \neq p(j) ),$

\medskip

\noindent then it remains injective after the update, 
i.e., check whether

\medskip 

\noindent ${\sf (Inj)}(p) \wedge {\sf Update}(p, p', a, a') \wedge 
(1 \leq c \leq m \wedge 1 \leq d \leq m \wedge c \neq d \wedge p'(c) = p'(d)) \models\bot$. 

\medskip

We need to deal with alternations of quantifiers in the extension.
The extension formulas $\K$ are augmented clauses of the form

\smallskip
~~~~~~$\forall x_1,...,x_n.\;(\Phi(x_1, \dots, x_n) \vee C(x_1, \dots, x_n)),$

\smallskip 
where $\Phi(x_1, \dots, x_n)$ is an {\em arbitrary first-order 
formula} in the base signature with free variables $x_1, \dots, x_n$ 
and $C(x_1, \dots, x_n)$ is a {\em clause} in the extended signature.
As input for H-PILoT,
extended clauses may be either written as
$\forall \x.\;(\Phi(\x) \vee C(\x))$ or as
$\forall \x.\;(\Phi(\x) \rightarrow C'(\x))$.
The input file for \hpilot  looks as follows.

\begin{MyFVerbatim}
% Updating of priorities of processes
% File update_AE.loc
Base_functions:={(+,2), (-, 2)}
Extension_functions:={(a', 1, 2, bool), (a, 1, 1, bool), (p', 1, 2, real), (p, 1, 1, real)}
Relations:={(<=, 2), (<, 2), (>, 2)}
Constants:={(x, real), (y, real)}

% K
Clauses := 
 (FORALL i). _1 <= i, i <= m --> _0 <= p(i);
 (FORALL i). { AND(_1 <= i, i <= m, 
    (FORALL j). (AND(_1 <= j, j <= m, NOT(j = i))
                                        --> p(i) > p(j)))}
                                            --> a'(i) = _1;

 (FORALL i). { AND(_1 <= i, i <= m, 
    (FORALL j). (AND(_1 <= j, j <= m, NOT(j = i)) 
                                        --> p(i) > p(j)))} 
                                            --> p'(i) = x;
 (FORALL i). { AND(_1 <= i, i <= m, 
    NOT((FORALL j,i).(AND(_1 <= j, j <= m, NOT(j = i)) 
                                        --> p(i) > p(j))))} 
                                            --> a'(i) = _0;
 (FORALL i). { AND(_1 <= i, i <= m, 
    NOT((FORALL j).(AND(_1 <= j, j <= m, NOT(j = i)) 
                                        --> p(i) > p(j))))} 
                                            --> p'(i) = p(i) + y;
         
 (FORALL i,j). _1 <= i, i <= m, _1 <= j, j <= m, p(i) = p(j) 
                                            --> i = j;

Query := _1 <= c;
         c <= m;
         _1 <= d;
         d <= m;
         x <= _0;
         y > _0;
         NOT(c=d);
         p'(c) = p'(d);
\end{MyFVerbatim}

\noindent
The curly braces '$\{$', '$\}$' are required to mark the beginning and the end of the base formula $\Phi$.

\section{System evaluation}

We have used H-PILoT on a variety of local extensions and on chains of local 
extensions.
An overview of the tests we made 
is given below.
For these tests, we have used Yices as the back-end solver for H-PILoT.
We distinguish between satisfiable and unsatisfiable problems.

\bigskip
\noindent{\bf Unsatisfiable Problems.} 
For simple unsatisfiable problems, there hardly is any difference 
in run-time whether one uses H-PILoT ~or an 
SMT-solver directly. This is due to the fact 
that a good SMT-solver uses the heuristic of trying out all the 
occurring ground terms as instantiations of universal quantifiers.
For local extensions this is always sufficient to derive a contradiction.

When we consider chains of extensions the picture changes
dramatically.  On one test example -- the array insertion of Section
\ref{sec-examples} which used a chain of two local extensions -- Yices
performed considerably slower than H-PILoT: The original problem took
Yices over 5 minutes to solve. The hierarchical reduction yielded 113
clauses of the background theory (integers) which were proved to be
unsatisfiable by Yices in a mere 0.07s.

\bigskip
\noindent{\bf Satisfiable Problems.}
For satisfiable problems over local theory extensions, 
H-PILoT always provides the right answer. In local extensions, 
H-PILoT is a decision procedure whereas completeness of other 
SMT-solvers is not guaranteed.  
In the test runs, Yices either ran out of memory or took more than 
6 hours when given any of the unreduced problems. 
This even was the case for small problems, 
e.g., problems over the reals with less than ten clauses.
With H-PILoT ~as a front end, Yices solved all the satisfiable problems in less than a second
with the single exception of monotone functions over posets/distributive lattices.

\subsection{Test runs for H-PILoT}\label{tests}

We analyzed the following examples.
The satisfiable variant of a problem carries the suffix ``.sat''.

\noindent 
\begin{description}
\item[{\sf array insert}.]   Insertion of an element into a sorted integer array.
                           This is the example from Section \ref{sec-examples}.
                                          Arrays are definitional extensions here.

\item[{\sf array insert ($\exists$).}]   Insertion of an element into a sorted integer array.
  Arrays are definitional extensions here.
   Alternate version with (implicit)
existential quantifier.
 \item[{\sf array insert  (linear).}]   Linear version of {\sf array insert}.
 \item[{\sf array insert real.}]               Like {\sf array insert } but with an array of reals.
 \item[{\sf array insert real (linear).}]      Linear version of {\sf array insert real}.
\item[{\sf update process priorities ($\forall\exists$).}]                    Updating of priorities of processes.
         This is the example from Section \ref{sec-extended-locality}. We have an $\forall\exists$-clause.
\item[{\sf list1.}]                             Made-up example of integer lists. 
                                          Some arithmetic is required  
 \item[{\sf chain1.}]                            Simple test for chains of extensions (plus transitivity). 
\item[{\sf chain2.}]                             Simple test for chains of extensions (plus transitivity and arithmetic).
\item[{\sf double array insert.}]      A sorted array is updated twice. 
This is the example from Section~\ref{sec-arrays}. 
It is inside the array property fragment.
\item[{\sf mono.}]                               Two monotone functions over integers/reals for SMT solver.
\item[{\sf mono for distributive lattices.R.}]    Two monotone functions over a distributive lattice. The axioms for a 
                                        distributive lattice are stated together with the definition of a relation $R$: 
                                         $R(x, y) :\Leftrightarrow x \wedge y = x$. Monotonicity  of $f$ (respectively of $g$) is given in 
                                         terms of $R$:  $R(x, y) \rightarrow R(f(x), f(y))$.  Flag \verb+-freeType+ must be used. 
\item[{\sf mono for distributive lattices.}]     Same as {\sf mono for distributive lattices.R} except that no relation 
 $R$ is defined. Monotonicity of the two functions $f, g$ is directly given:
                                         $x \wedge y = x\rightarrow f(x) \wedge
f(y) = f(x)$. 
                                         Flag \verb+-freeType+ must be used. 
 \item[{\sf mono for poset.}]                    Two monotone functions over a poset with poset axioms  as in Section \ref{sec-examples}. Same as {\sf mono}, 
                                         except the order is modeled by a relation $R$. 
  \item[{\sf mono for total order.}]             Same as {\sf mono} except linearity is an axiom.
                                         This makes no difference unless SPASS is used.
\item[{\sf own.}]                               Simple test for monotone function.
\phantom{(Bradley/Manna)} 
   \item[{\sf mvLogic/mv1.}]                   The example for MV-algebras from Section \ref{sec-mv}. The {\L}ukasiewicz connectives can be defined 
 in terms of the (real) operations 
$+,-,\leq$. Linearity is deducible from axioms.
   \item[{\sf mvLogic/mv2.}]                       Example for MV-algebras. The {\L}ukasiewicz connectives can be
defined 
   in terms of 
$+,-,\leq$. 
  \item[{\sf mvLogic/bl1.}]                       Example for MV-algebras with BL axiom (redundantly) included. 
                                            The {\L}ukasiewicz connectives can be defined in terms of $+,-,\leq$.  
   \item[{\sf mvLogic/example\_6.1.}]              Example for MV-algebras with monotone and bounded function.
                                            The {\L}ukasiewicz connectives can be defined in terms of $+,-,\leq$. 
   \item[{\sf RBC\_simple.}]                       Example with train controller. 
   \item[{\sf RBC\_variable2.}]                    Example with train controller. 
\end{description}

\subsection{Test results}

The running times are given in User + sys times (in s). Run on an
Intel Xeon 3 GHz, 512 kB cache; median of 100 runs (entries
marked with $^1$: 10 runs; marked with $^2$: 3 runs).  The third
column lists the number of clauses produced; ``unknown'' means Yices
answer was \verb+unknown+, ``out. mem.'' means out of memory and time
out was set at 6h.  Yices version 1.0.19 was used.

The answer ``unknown$^*$'' for the satisfiable examples with monotone
functions over distributive lattices/posets (H-PILoT followed by
Yices) is due to the fact that Yices cannot handle the universal
axioms of distributive lattices/posets.  A translation of such
problems to SAT provides a decision procedure
(cf. \cite{sofronie-cade05} and also \cite{sofronie-jsc03}).

\bigskip

\noindent
{\footnotesize
\begin{center}
$\mbox{\hspace{0cm}}${\begin{tabular}{|l|l|l|r@{.}l|r@{.}l|r@{.}l|}
\hline
{\bf Name} & {\bf \ status} &{\bf \#{}cl.}& \multicolumn{2}{|c|}{{\bf H-PILoT}} & \multicolumn{2}{|c|}{{\bf H-PILoT}} & \multicolumn{2}{|c|}{{\bf  yices}} \\ 
           &                &             & \multicolumn{2}{|c|}{{\bf + yices}} & \multicolumn{2}{|c|}{{\bf + yices}} & \multicolumn{2}{|c|}{} \\
           &                &             & \multicolumn{2}{|c|}{}              & \multicolumn{2}{|c|}{stop at $\K [G]$}  & \multicolumn{2}{|c|}{} \\
\hline
array insert (implicit $\exists$)     & Unsat & 310 & 0&29 &  0&06 & 0&36                           \\ \hline
array insert (implicit $\exists$).sat & Sat   & 196 & 0&13 &  0&04 & \multicolumn{2}{|c|}{time out}    \\ \hline
array insert                          & Unsat & 113 & 0&07 &  0&03 & 318&22$^1$             \\ \hline
array insert (linear version)         & Unsat & 113 & 0&07 &  0&03 & 7970&53$^2$        \\ \hline
array insert.sat                      & Sat   & 111 & 0&07 &  0&03 & \multicolumn{2}{|c|}{time out}   \\ \hline
array insert real                     & Unsat & 113 & 0&07 &  0&03 & 360&00$^1$              \\ \hline
array insert real (linear)            & Unsat & 113 & 0&07 &  0&03 & 7930&00$^2$              \\ \hline
array insert real.sat                 & Sat   & 111 & 0&07 &  0&03 & \multicolumn{2}{|c|}{time out}   \\ \hline
 update process priorities            & Unsat & 45  & 0&02 &  0&02 & 0&03                         \\ \hline
 update process priorities.sat        & Sat   & 37  & 0&02 &  0&02 & \multicolumn{2}{|c|}{unknown}   \\ \hline
list1                                 & Unsat & 18  & 0&02 &  0&01 & 0&02                          \\ \hline
list1.sat                             & Sat   & 16  & 0&02 &  0&01 & \multicolumn{2}{|c|}{unknown}   \\ \hline
chain1                                & Unsat & 22  & 0&01 &  0&01 & 0&02                 \\ \hline
chain2                                & Unsat & 46  & 0&02 &  0&02 & 0&02                \\ \hline
mono                                  & Unsat & 20  & 0&01 &  0&01 & 0&01   \\ \hline
mono.sat                              & Sat   & 20  & 0&01 & 0&01 & \multicolumn{2}{|c|}{unknown}  \\\hline 
mono for distributive lattices.R & Unsat & 27  & 0&22 &  0&06 &  0&03                         \\ \hline
mono for distributive lattices.R.sat & Sat   & 26  & \multicolumn{2}{|c|}{unknown$^*$} & \multicolumn{2}{|c|}{unknown$^*$} &
\multicolumn{2}{|c|}{unknown}          \\ \hline
mono for distributive lattices   & Unsat & 17  & 0&01 &  0&01 &  0&02                          \\ \hline
mono for distributive lattices.sat & Sat   & 17  & 0&01 &  0&01 &
\multicolumn{2}{|c|}{unknown}   \\ \hline
mono for poset                        & Unsat & 20  & 0&02 &  0&02 &  0&02          
               \\ \hline
mono for poset.sat               & Sat   & 19  & \multicolumn{2}{|c|}{unknown$^*$}
& \multicolumn{2}{|c|}{unknown$^*$} & \multicolumn{2}{|c|}{unknown}   \\ \hline
mono for total order                  & Unsat & 20  & 0&02 &  0&02 &  0&02          
               \\ \hline     
own                                   & Unsat & 16  & 0&01 &  0&01 &  0&01          
               \\ \hline
mvLogic/mv1                           & Unsat & 10  & 0&01 &  0&01 &  0&02          
               \\ \hline
mvLogic/mv1.sat                       & Sat   & 8   & 0&01 &  0&01 & 
\multicolumn{2}{|c|}{unknown}  \\ \hline
mvLogic/mv2                           & Unsat & 8   & 0&01 &  0&01 &  0&06  \\ \hline
mvLogic/bl1                           & Unsat & 22  & 0&02 &  0&01 &  0&03          
               \\ \hline
mvLogic/example\_6.1                  & Unsat & 10  & 0&01 &  0&01 &  0&03          
               \\ \hline
mvLogic/example\_6.1.sat              & Sat   & 10  & 0&01 &  0&01 &
\multicolumn{2}{|c|}{unknown}   \\ \hline
RBC\_simple                           & Unsat & 42  & 0&03 &  0&02 &  0&03          
               \\ \hline
double array insert                   & Unsat & 791 & 1&16 &  0&20 &  0&07          
               \\ \hline
double array insert                   & Sat   & 790 & 1&10 &  0&20 & 
\multicolumn{2}{|c|}{unknown}   \\ \hline
RBC\_simple.sat                       & Sat   & 40  & 0&03 &  0&02 & \multicolumn{2}{|c|}{out. mem.} \\ \hline
RBC\_variable2                        & Unsat & 137 & 0&08 &  0&04 &  0&04          
               \\ \hline
RBC\_variable2.sat                    & Sat   & 136 & 0&08 &  0&04 & \multicolumn{2}{|c|}{out. mem.}    \\ \hline
\end{tabular}}
\end{center}
}

\medskip
\section{Examples}

\subsection{A case study}\label{sec-case-study}

In \cite{sofronie-ihlemann-jacobs:tacas} we used 
H-PILoT for verifying the correctness of the controller of a 
system of trains moving on a linear track.
In \cite{faber-ihlemann-jacobs-sofronie-ifm10}, 
H-PILoT's ability to decide the  
pointer fragment of Section~\ref{sec-pointers} has been used
in the verification of real-time systems which exhibit rich and 
dynamic data structures.
There H-PILoT was part of a  tool chain employed for the 
verification of a case study
from the European Train Control System standard, describing the controller 
of a system of trains moving on a rail track with complex topology -- modeled 
using two-sorted pointer structures.  
The tool chain received as input a 
high level specification and a formula (a safety property), 
and generated proof obligations, which were 
automatically verified using H-PILoT (with Yices).

The verification problem we considered are expressed as 
satisfiability problems for universally quantified formulas, hence
cannot be solved by SMT-solvers alone.
The experimental results show H-PILoT to be a very efficient tool for 
the discharging of all the proof tasks of the case study.
The full type system implemented in \hpilot increased the efficiency 
considerably by blocking unnecessary instantiations.
The tool chain used in the case study range from a specification 
language for real-time systems 
called COD to the translation of such a specification via 
phase-event automata (Syspect/PEA)
to transition constraint systems (TCS) which can then be exported to H-PILoT.
The invariant for every transition in the TCS was checked.

Since the invariant was too
complex to be handled by the clausifier of \hpilot we checked the 
invariant for every transition in two parts yielding 92 proof obligations.  
Further, we performed 
tests to ensure that the specifications are consistent.
The time to compute the TCS from the specification 
was insignificant. The overall time to verify all transition updates with
Yices and \hpilot in the unsatisfiable case (when the invariant is correct)
does not differ much. There is one example -- the speed update -- on which 
\hpilot was 5 times faster than Yices alone.

We also made tests with the verification of a set of conditions 
which was not inductive over all transitions. 
Here, \hpilot was able to provide a model after 8s  
whereas Yices detected unsatisfiability for 17 problems, returned ``unknown'' 
for 28, and timed out once. 
For the consistency check \hpilot was able to provide a model after 3s, 
whereas Yices answered ``unknown''.  







During the development of the case study
\hpilot helped us finding the correct transition invariants by providing models
for satisfiable sets of clauses (occurring when the safety formulae were not 
invariant under transitions).

\subsection{A run example of H-PILoT}
\label{sec-run}

We consider an example taken from \cite{bradley:book} (Example~11.10).
The input file looks as follows.

\begin{MyFVerbatim}
% arrays_from_book.loc
Base_functions:={(+, 2), (-, 2)}
Extension_functions:={(a, 1, 1, int, int), (b, 1, 1, int, int)}
Relations:={(<=, 2)}

% K
Formulas :=  
 AND( (FORALL i). (AND(l <= i, i <= u) --> a(i) = b(i)),
       NOT((FORALL i). (AND(l <= i, i <= u + _1) --> 
                         write(a, u + _1, b(u +_1))(i) = b(i))));
\end{MyFVerbatim}

\noindent The arrays $a$ and $b$ are considered to be equal between 
the constants $l$ and $u$. We prove that if we update $a$ at $u+1$ to 
$b(u+1)$ then $a$ and $b$ should be equal between $l$ 
and $u+1$. The formula above denies this and should therefore be 
inconsistent. We call H-PILoT with

\medskip

\verb+hpilot.opt -preprocess -prClauses arrays_from_book.loc+

\medskip

\noindent \verb+-preprocess+ is needed as usual; we use \verb+-prClauses+ to get a trace of
the program.
(Because the array keyword \verb+write+ appears in the input we don't have to use the
flag \verb+-arrays+:
it is implicit.) The trace looks as follows 
(to improve readability we aligned the level labels
and often left out the listing of the extension ground terms 
due to space constraints).

First, H-PILoT reads the input and clausifies the formula.

\begin{MySmallVerbatim}
 ********************************************** Starting hpilot**********************************************
 arrays_from_book.loc
 Adding formula: 
 AND( (FORALL i). (AND( l <= i, i <= u) --> a(i) = b(i)), 
      NOT((FORALL i). (AND( l <= i, i <= +(u, _1)) --> read(write(a, +(u, _1),b(+(u, _1))), i) = b(i))))
 done.
 Clausifying formulas...
 (FORALL z_1). OR( NOT(l <= z_1), NOT(z_1 <= u), a(z_1) = b(z_1))
 l <= sk_1
 sk_1 <= +(u, _1)
 NOT(read(write(a, +(u, _1), b(+(u, _1))), sk_1) = b(sk_1))
 Yielding 4 new clauses: 
 read(write(a, +(u, _1), b(+(u, _1))), sk_1) = b(sk_1) --->    L: 0; Extension ground terms: b(sk_1), b(+(u, _1))
  ---> sk_1 <= +(u, _1)                                        L: 0; Extension ground terms: 
  ---> l <= sk_1                                               L: 0; Extension ground terms: 
 [z_1] l <= z_1, z_1 <= u ---> a(z_1) = b(z_1)                 L: 0; Extension ground terms: 
\end{MySmallVerbatim}

H-PILoT then replaces array writes by introducing new arrays:\\ 
${\rm write}(a,u + 1,b(u + 1))$ is replaced by:
  $\forall i. i \neq u + 1 \rightarrow a_{w1}(i) = a(i)$ and  $a_{w1}(u+1) = b(u+1)$. 
 To remain in the decidable fragment, \hpilot replaces $\forall i. i
\neq u + 1 \rightarrow a_{w1}(i) =a(i)$ with  
$\forall i. i \leq u + 1 - 1 \rightarrow a_{w1}(i) = a(i)$ and
$\forall i. u + 1 + 1 \leq i \rightarrow a_{w1}(i) = a(i)$.

\begin{MySmallVerbatim}
 Replacing writes...
 We have 1 levels.
 Our base theory is: 
 empty.
 Splitting clause [i]  ---> i = +(u, _1), a_w1(i) = a(i)                      L: 1; 
terms:  on eq i = +(u, _1)
 Checking APF for clause [i] i <= -(+(u, _1), _1) ---> a_w1(i) = a(i)         L: 0;
Extension ground terms: ---> true
 Checking APF for clause [i] +(+(u, _1), _1) <= i ---> a_w1(i) = a(i)         L: 0;
Extension ground terms: ---> true
 Checking APF for clause [z_1] l <= z_1, z_1 <= u ---> a(z_1) = b(z_1)        L: 1;
Extension ground terms: ---> true
 Recalculating all levels.
\end{MySmallVerbatim}

\noindent
\hpilot then flattens and linearizes the result.

\begin{MySmallVerbatim}
After rewriting we have as clauses K: 
 [i, x_1] x_1 = i, i <= -(+(u, _1), _1) ---> a_w1(i) = a(x_1)  L: 1; Extension ground terms: 
 [i, x_1] x_1 = i, +(+(u, _1), _1) <= i ---> a_w1(i) = a(x_1)  L: 1; Extension ground terms: 
 [z_1, x_1] x_1 = z_1, l <= z_1, z_1 <= u ---> a(z_1) = b(x_1) L: 1; Extension ground terms: 
 and as query: 
  ---> l <= sk_1                                               L: 0; Extension ground terms: 
  ---> sk_1 <= +(u, _1)                                        L: 0; Extension ground terms: 
 a_w1(sk_1) = b(sk_1) --->                 L: 1; Extension ground terms: a_w1(sk_1), b(sk_1)
  ---> a_w1(+(u, _1)) = b(+(u, _1))  L: 1; Extension ground terms: a_w1(+(u, _1)),b(+(u, _1))
 Our query G is : 
  ---> l <= sk_1                                               L: 0; Extension ground terms: 
  ---> sk_1 <= +(u, _1)                                        L: 0; Extension ground terms: 
 a_w1(sk_1) = b(sk_1) --->                 L: 1; Extension ground terms: a_w1(sk_1), b(sk_1)
  ---> a_w1(+(u, _1)) = b(+(u, _1))  L: 1; Extension ground terms: a_w1(+(u, _1)),b(+(u, _1))
 xxxxxxxxxxx End preprocessing.
\end{MySmallVerbatim}

\noindent
H-PILoT then calculates the set of instances $\K[\Psi(G)]$
and simplifies the terms to avoid redundant instances of clauses (e.g.\ $u {+} 1 {-}1$ is replaced by $u$).

\begin{MySmallVerbatim}
We have 5 index terms for minimal locality l, sk_1, u, +(u, _1), +(u, _2)
K has 3 members.
[i, x_1] x_1 = i, i <= -(+(u, _1), _1) ---> a_w1(i) = a(x_1)                          L: 1; 
[i, x_1] x_1 = i, +(+(u, _1), _1) <= i ---> a_w1(i) = a(x_1)                          L: 1; 
[z_1, x_1] x_1 = z_1, l <= z_1, z_1 <= u ---> a(z_1) = b(x_1)                         L: 1; 
Computing K<G>... 
K<G> looks as follows: 
K_G has 75 members.
[] l = l, l <= -(+(u, _1), _1) ---> a_w1(l) = a(l)                                    L: 0; 
[] l = sk_1, sk_1 <= -(+(u, _1), _1) ---> a_w1(sk_1) = a(l)                           L: 0; 
[] l = u, u <= -(+(u, _1), _1) ---> a_w1(u) = a(l)                                    L: 0; 
[] l = +(u, _1), +(u, _1) <= -(+(u, _1), _1) ---> a_w1(+(u, _1)) = a(l)               L: 0; 
[] l = +(u, _2), +(u, _2) <= -(+(u, _1), _1) ---> a_w1(+(u, _2)) = a(l)               L: 0; 
[] sk_1 = l, l <= -(+(u, _1), _1) ---> a_w1(l) = a(sk_1)                              L: 0; 
[] sk_1 = sk_1, sk_1 <= -(+(u, _1), _1) ---> a_w1(sk_1) = a(sk_1)                     L: 0; 
[] sk_1 = u, u <= -(+(u, _1), _1) ---> a_w1(u) = a(sk_1)                              L: 0; 
[] sk_1 = +(u, _1), +(u, _1) <= -(+(u, _1), _1) ---> a_w1(+(u, _1)) = a(sk_1)         L: 0; 
[] sk_1 = +(u, _2), +(u, _2) <= -(+(u, _1), _1) ---> a_w1(+(u, _2)) = a(sk_1)         L: 0; 
[] u = l, l <= -(+(u, _1), _1) ---> a_w1(l) = a(u)                                    L: 0; 
[] u = sk_1, sk_1 <= -(+(u, _1), _1) ---> a_w1(sk_1) = a(u)                           L: 0; 
[] u = u, u <= -(+(u, _1), _1) ---> a_w1(u) = a(u)                                    L: 0; 
[] u = +(u, _1), +(u, _1) <= -(+(u, _1), _1) ---> a_w1(+(u, _1)) = a(u)               L: 0; 
[] u = +(u, _2), +(u, _2) <= -(+(u, _1), _1) ---> a_w1(+(u, _2)) = a(u)               L: 0; 
[] +(u, _1) = l, l <= -(+(u, _1), _1) ---> a_w1(l) = a(+(u, _1))                      L: 0; 
[] +(u, _1) = sk_1, sk_1 <= -(+(u, _1), _1) ---> a_w1(sk_1) = a(+(u, _1))             L: 0; 
[] +(u, _1) = u, u <= -(+(u, _1), _1) ---> a_w1(u) = a(+(u, _1))                      L: 0; 
[] +(u, _1) = +(u, _1), +(u, _1) <= -(+(u, _1), _1) ---> a_w1(+(u, _1)) = a(+(u,_1))  L: 0; 
[] +(u, _1) = +(u, _2), +(u, _2) <= -(+(u, _1), _1) ---> a_w1(+(u, _2)) = a(+(u,_1))  L: 0; 
[] +(u, _2) = l, l <= -(+(u, _1), _1) ---> a_w1(l) = a(+(u, _2))                      L: 0; 
[] +(u, _2) = sk_1, sk_1 <= -(+(u, _1), _1) ---> a_w1(sk_1) = a(+(u, _2))             L: 0; 
[] +(u, _2) = u, u <= -(+(u, _1), _1) ---> a_w1(u) = a(+(u, _2))                      L: 0; 
[] +(u, _2) = +(u, _1), +(u, _1) <= -(+(u, _1), _1) ---> a_w1(+(u, _1)) = a(+(u,_2))  L: 0; 
[] +(u, _2) = +(u, _2), +(u, _2) <= -(+(u, _1), _1) ---> a_w1(+(u, _2)) = a(+(u,_2))  L: 0; 
[] l = l, +(+(u, _1), _1) <= l ---> a_w1(l) = a(l)                                    L: 0; 
[] l = sk_1, +(+(u, _1), _1) <= sk_1 ---> a_w1(sk_1) = a(l)                           L: 0; 
[] l = u, +(+(u, _1), _1) <= u ---> a_w1(u) = a(l)                                    L: 0; 
[] l = +(u, _1), +(+(u, _1), _1) <= +(u, _1) ---> a_w1(+(u, _1)) = a(l)               L: 0; 
[] l = +(u, _2), +(+(u, _1), _1) <= +(u, _2) ---> a_w1(+(u, _2)) = a(l)               L: 0; 
[] sk_1 = l, +(+(u, _1), _1) <= l ---> a_w1(l) = a(sk_1)                              L: 0; 
[] sk_1 = sk_1, +(+(u, _1), _1) <= sk_1 ---> a_w1(sk_1) = a(sk_1)                     L: 0; 
[] sk_1 = u, +(+(u, _1), _1) <= u ---> a_w1(u) = a(sk_1)                              L: 0; 
[] sk_1 = +(u, _1), +(+(u, _1), _1) <= +(u, _1) ---> a_w1(+(u, _1)) = a(sk_1)         L: 0; 
[] sk_1 = +(u, _2), +(+(u, _1), _1) <= +(u, _2) ---> a_w1(+(u, _2)) = a(sk_1)         L: 0; 
[] u = l, +(+(u, _1), _1) <= l ---> a_w1(l) = a(u)                                    L: 0; 
[] u = sk_1, +(+(u, _1), _1) <= sk_1 ---> a_w1(sk_1) = a(u)                           L: 0; 
[] u = u, +(+(u, _1), _1) <= u ---> a_w1(u) = a(u)                                    L: 0; 
[] u = +(u, _1), +(+(u, _1), _1) <= +(u, _1) ---> a_w1(+(u, _1)) = a(u)               L: 0; 
[] u = +(u, _2), +(+(u, _1), _1) <= +(u, _2) ---> a_w1(+(u, _2)) = a(u)               L: 0; 
[] +(u, _1) = l, +(+(u, _1), _1) <= l ---> a_w1(l) = a(+(u, _1))                      L: 0; 
[] +(u, _1) = sk_1, +(+(u, _1), _1) <= sk_1 ---> a_w1(sk_1) = a(+(u, _1))             L: 0; 
[] +(u, _1) = u, +(+(u, _1), _1) <= u ---> a_w1(u) = a(+(u, _1))                      L: 0; 
[] +(u, _1) = +(u, _1), +(+(u, _1), _1) <= +(u, _1) ---> a_w1(+(u, _1)) = a(+(u,_1))  L: 0; 
[] +(u, _1) = +(u, _2), +(+(u, _1), _1) <= +(u, _2) ---> a_w1(+(u, _2)) = a(+(u,_1))  L: 0; 
[] +(u, _2) = l, +(+(u, _1), _1) <= l ---> a_w1(l) = a(+(u, _2))                      L: 0; 
[] +(u, _2) = sk_1, +(+(u, _1), _1) <= sk_1 ---> a_w1(sk_1) = a(+(u, _2))             L: 0; 
[] +(u, _2) = u, +(+(u, _1), _1) <= u ---> a_w1(u) = a(+(u, _2))                      L: 0; 
[] +(u, _2) = +(u, _1), +(+(u, _1), _1) <= +(u, _1) ---> a_w1(+(u, _1)) = a(+(u,_2))  L: 0; 
[] +(u, _2) = +(u, _2), +(+(u, _1), _1) <= +(u, _2) ---> a_w1(+(u, _2)) = a(+(u,_2))  L: 0; 
[] l = l, l <= l, l <= u ---> a(l) = b(l)                                             L: 0; 
[] sk_1 = l, l <= l, l <= u ---> a(l) = b(sk_1)                                       L: 0; 
[] u = l, l <= l, l <= u ---> a(l) = b(u)                                             L: 0; 
[] +(u, _1) = l, l <= l, l <= u ---> a(l) = b(+(u, _1))                               L: 0; 
[] +(u, _2) = l, l <= l, l <= u ---> a(l) = b(+(u, _2))                               L: 0; 
[] l = sk_1, l <= sk_1, sk_1 <= u ---> a(sk_1) = b(l)                                 L: 0; 
[] sk_1 = sk_1, l <= sk_1, sk_1 <= u ---> a(sk_1) = b(sk_1)                           L: 0; 
[] u = sk_1, l <= sk_1, sk_1 <= u ---> a(sk_1) = b(u)                                 L: 0; 
[] +(u, _1) = sk_1, l <= sk_1, sk_1 <= u ---> a(sk_1) = b(+(u, _1))                   L: 0; 
[] +(u, _2) = sk_1, l <= sk_1, sk_1 <= u ---> a(sk_1) = b(+(u, _2))                   L: 0; 
[] l = u, l <= u, u <= u ---> a(u) = b(l)                                             L: 0; 
[] sk_1 = u, l <= u, u <= u ---> a(u) = b(sk_1)                                       L: 0; 
[] u = u, l <= u, u <= u ---> a(u) = b(u)                                             L: 0; 
[] +(u, _1) = u, l <= u, u <= u ---> a(u) = b(+(u, _1))                               L: 0; 
[] +(u, _2) = u, l <= u, u <= u ---> a(u) = b(+(u, _2))                               L: 0; 
[] l = +(u, _1), l <= +(u, _1), +(u, _1) <= u ---> a(+(u, _1)) = b(l)                 L: 0; 
[] sk_1 = +(u, _1), l <= +(u, _1), +(u, _1) <= u ---> a(+(u, _1)) = b(sk_1)           L: 0; 
[] u = +(u, _1), l <= +(u, _1), +(u, _1) <= u ---> a(+(u, _1)) = b(u)                 L: 0; 
[] +(u, _1) = +(u, _1), l <= +(u, _1), +(u, _1) <= u ---> a(+(u, _1)) = b(+(u,_1))    L: 0; 
[] +(u, _2) = +(u, _1), l <= +(u, _1), +(u, _1) <= u ---> a(+(u, _1)) = b(+(u,_2))    L: 0; 
[] l = +(u, _2), l <= +(u, _2), +(u, _2) <= u ---> a(+(u, _2)) = b(l)                 L: 0; 
[] sk_1 = +(u, _2), l <= +(u, _2), +(u, _2) <= u ---> a(+(u, _2)) = b(sk_1)           L: 0; 
[] u = +(u, _2), l <= +(u, _2), +(u, _2) <= u ---> a(+(u, _2)) = b(u)                 L: 0; 
[] +(u, _1) = +(u, _2), l <= +(u, _2), +(u, _2) <= u ---> a(+(u, _2)) = b(+(u,_1))    L: 0; 
[] +(u, _2) = +(u, _2), l <= +(u, _2), +(u, _2) <= u ---> a(+(u, _2)) = b(+(u,_2))    L: 0; 
\end{MySmallVerbatim}

\noindent The result is then purified:

\begin{MySmallVerbatim}
computing defs ...
 We have the following definitions: 
  ---> e_1 = a(l)                                                        L: 0; Extension ground terms: a(l)
  ---> e_2 = a(sk_1)                                                     L: 0; Extension ground terms: a(sk_1)
  ---> e_3 = a(u)                                                        L: 0; Extension ground terms: a(u)
  ---> e_4 = a(+(u, _1))                                                 L: 0; Extension ground terms: a(+(u, _1))
  ---> e_5 = a(+(u, _2))                                                 L: 0; Extension ground terms: a(+(u, _2))
  ---> e_6 = a_w1(l)                                                     L: 0; Extension ground terms: a_w1(l)
  ---> e_7 = a_w1(sk_1)                                                  L: 0; Extension ground terms: a_w1(sk_1)
  ---> e_8 = a_w1(u)                                                     L: 0; Extension ground terms: a_w1(u)
  ---> e_9 = a_w1(+(u, _1))                                              L: 0; Extension ground terms: a_w1(+(u, _1))
  ---> e_10 = a_w1(+(u, _2))                                             L: 0; Extension ground terms: a_w1(+(u, _2))
  ---> e_11 = b(l)                                                       L: 0; Extension ground terms: b(l)
  ---> e_12 = b(sk_1)                                                    L: 0; Extension ground terms: b(sk_1)
  ---> e_13 = b(u)                                                       L: 0; Extension ground terms: b(u)
  ---> e_14 = b(+(u, _1))                                                L: 0; Extension ground terms: b(+(u, _1))
  ---> e_15 = b(+(u, _2))                                                L: 0; Extension ground terms: b(+(u, _2))
 Purified: 
 K_G has 75 members.
 [] l = l, l <= -(+(u, _1), _1) ---> e_6 = e_1                           L: 0; Extension ground terms: 
 [] l = sk_1, sk_1 <= -(+(u, _1), _1) ---> e_7 = e_1                     L: 0; Extension ground terms: 
 [] l = u, u <= -(+(u, _1), _1) ---> e_8 = e_1                           L: 0; Extension ground terms: 
 [] l = +(u, _1), +(u, _1) <= -(+(u, _1), _1) ---> e_9 = e_1             L: 0; Extension ground terms: 
 [] l = +(u, _2), +(u, _2) <= -(+(u, _1), _1) ---> e_10 = e_1            L: 0; Extension ground terms: 
 [] sk_1 = l, l <= -(+(u, _1), _1) ---> e_6 = e_2                        L: 0; Extension ground terms: 
 [] sk_1 = sk_1, sk_1 <= -(+(u, _1), _1) ---> e_7 = e_2                  L: 0; Extension ground terms: 
 [] sk_1 = u, u <= -(+(u, _1), _1) ---> e_8 = e_2                        L: 0; Extension ground terms: 
 [] sk_1 = +(u, _1), +(u, _1) <= -(+(u, _1), _1) ---> e_9 = e_2          L: 0; Extension ground terms: 
 [] sk_1 = +(u, _2), +(u, _2) <= -(+(u, _1), _1) ---> e_10 = e_2         L: 0; Extension ground terms: 
 [] u = l, l <= -(+(u, _1), _1) ---> e_6 = e_3                           L: 0; Extension ground terms: 
 [] u = sk_1, sk_1 <= -(+(u, _1), _1) ---> e_7 = e_3                     L: 0; Extension ground terms: 
 [] u = u, u <= -(+(u, _1), _1) ---> e_8 = e_3                           L: 0; Extension ground terms: 
 [] u = +(u, _1), +(u, _1) <= -(+(u, _1), _1) ---> e_9 = e_3             L: 0; Extension ground terms: 
 [] u = +(u, _2), +(u, _2) <= -(+(u, _1), _1) ---> e_10 = e_3            L: 0; Extension ground terms: 
 [] +(u, _1) = l, l <= -(+(u, _1), _1) ---> e_6 = e_4                    L: 0; Extension ground terms: 
 [] +(u, _1) = sk_1, sk_1 <= -(+(u, _1), _1) ---> e_7 = e_4              L: 0; Extension ground terms: 
 [] +(u, _1) = u, u <= -(+(u, _1), _1) ---> e_8 = e_4                    L: 0; Extension ground terms: 
 [] +(u, _1) = +(u, _1), +(u, _1) <= -(+(u, _1), _1) ---> e_9 = e_4      L: 0; Extension ground terms: 
 [] +(u, _1) = +(u, _2), +(u, _2) <= -(+(u, _1), _1) ---> e_10 = e_4     L: 0; Extension ground terms: 
 [] +(u, _2) = l, l <= -(+(u, _1), _1) ---> e_6 = e_5                    L: 0; Extension ground terms: 
 [] +(u, _2) = sk_1, sk_1 <= -(+(u, _1), _1) ---> e_7 = e_5              L: 0; Extension ground terms: 
 [] +(u, _2) = u, u <= -(+(u, _1), _1) ---> e_8 = e_5                    L: 0; Extension ground terms: 
 [] +(u, _2) = +(u, _1), +(u, _1) <= -(+(u, _1), _1) ---> e_9 = e_5      L: 0; Extension ground terms: 
 [] +(u, _2) = +(u, _2), +(u, _2) <= -(+(u, _1), _1) ---> e_10 = e_5     L: 0; Extension ground terms: 
 [] l = l, +(+(u, _1), _1) <= l ---> e_6 = e_1                           L: 0; Extension ground terms: 
 [] l = sk_1, +(+(u, _1), _1) <= sk_1 ---> e_7 = e_1                     L: 0; Extension ground terms: 
 [] l = u, +(+(u, _1), _1) <= u ---> e_8 = e_1                           L: 0; Extension ground terms: 
 [] l = +(u, _1), +(+(u, _1), _1) <= +(u, _1) ---> e_9 = e_1             L: 0; Extension ground terms: 
 [] l = +(u, _2), +(+(u, _1), _1) <= +(u, _2) ---> e_10 = e_1            L: 0; Extension ground terms: 
 [] sk_1 = l, +(+(u, _1), _1) <= l ---> e_6 = e_2                        L: 0; Extension ground terms: 
 [] sk_1 = sk_1, +(+(u, _1), _1) <= sk_1 ---> e_7 = e_2                  L: 0; Extension ground terms: 
 [] sk_1 = u, +(+(u, _1), _1) <= u ---> e_8 = e_2                        L: 0; Extension ground terms: 
 [] sk_1 = +(u, _1), +(+(u, _1), _1) <= +(u, _1) ---> e_9 = e_2          L: 0; Extension ground terms: 
 [] sk_1 = +(u, _2), +(+(u, _1), _1) <= +(u, _2) ---> e_10 = e_2         L: 0; Extension ground terms: 
 [] u = l, +(+(u, _1), _1) <= l ---> e_6 = e_3                           L: 0; Extension ground terms: 
 [] u = sk_1, +(+(u, _1), _1) <= sk_1 ---> e_7 = e_3                     L: 0; Extension ground terms: 
 [] u = u, +(+(u, _1), _1) <= u ---> e_8 = e_3                           L: 0; Extension ground terms: 
 [] u = +(u, _1), +(+(u, _1), _1) <= +(u, _1) ---> e_9 = e_3             L: 0; Extension ground terms: 
 [] u = +(u, _2), +(+(u, _1), _1) <= +(u, _2) ---> e_10 = e_3            L: 0; Extension ground terms: 
 [] +(u, _1) = l, +(+(u, _1), _1) <= l ---> e_6 = e_4                    L: 0; Extension ground terms: 
 [] +(u, _1) = sk_1, +(+(u, _1), _1) <= sk_1 ---> e_7 = e_4              L: 0; Extension ground terms: 
 [] +(u, _1) = u, +(+(u, _1), _1) <= u ---> e_8 = e_4                    L: 0; Extension ground terms: 
 [] +(u, _1) = +(u, _1), +(+(u, _1), _1) <= +(u, _1) ---> e_9 = e_4      L: 0; Extension ground terms: 
 [] +(u, _1) = +(u, _2), +(+(u, _1), _1) <= +(u, _2) ---> e_10 = e_4     L: 0; Extension ground terms: 
 [] +(u, _2) = l, +(+(u, _1), _1) <= l ---> e_6 = e_5                    L: 0; Extension ground terms: 
 [] +(u, _2) = sk_1, +(+(u, _1), _1) <= sk_1 ---> e_7 = e_5              L: 0; Extension ground terms: 
 [] +(u, _2) = u, +(+(u, _1), _1) <= u ---> e_8 = e_5                    L: 0; Extension ground terms: 
 [] +(u, _2) = +(u, _1), +(+(u, _1), _1) <= +(u, _1) ---> e_9 = e_5      L: 0; Extension ground terms: 
 [] +(u, _2) = +(u, _2), +(+(u, _1), _1) <= +(u, _2) ---> e_10 = e_5     L: 0; Extension ground terms: 
 [] l = l, l <= l, l <= u ---> e_1 = e_11                                L: 0; Extension ground terms: 
 [] sk_1 = l, l <= l, l <= u ---> e_1 = e_12                             L: 0; Extension ground terms: 
 [] u = l, l <= l, l <= u ---> e_1 = e_13                                L: 0; Extension ground terms: 
 [] +(u, _1) = l, l <= l, l <= u ---> e_1 = e_14                         L: 0; Extension ground terms: 
 [] +(u, _2) = l, l <= l, l <= u ---> e_1 = e_15                         L: 0; Extension ground terms: 
 [] l = sk_1, l <= sk_1, sk_1 <= u ---> e_2 = e_11                       L: 0; Extension ground terms: 
 [] sk_1 = sk_1, l <= sk_1, sk_1 <= u ---> e_2 = e_12                    L: 0; Extension ground terms: 
 [] u = sk_1, l <= sk_1, sk_1 <= u ---> e_2 = e_13                       L: 0; Extension ground terms: 
 [] +(u, _1) = sk_1, l <= sk_1, sk_1 <= u ---> e_2 = e_14                L: 0; Extension ground terms: 
 [] +(u, _2) = sk_1, l <= sk_1, sk_1 <= u ---> e_2 = e_15                L: 0; Extension ground terms: 
 [] l = u, l <= u, u <= u ---> e_3 = e_11                                L: 0; Extension ground terms: 
 [] sk_1 = u, l <= u, u <= u ---> e_3 = e_12                             L: 0; Extension ground terms: 
 [] u = u, l <= u, u <= u ---> e_3 = e_13                                L: 0; Extension ground terms: 
 [] +(u, _1) = u, l <= u, u <= u ---> e_3 = e_14                         L: 0; Extension ground terms: 
 [] +(u, _2) = u, l <= u, u <= u ---> e_3 = e_15                         L: 0; Extension ground terms: 
 [] l = +(u, _1), l <= +(u, _1), +(u, _1) <= u ---> e_4 = e_11           L: 0; Extension ground terms: 
 [] sk_1 = +(u, _1), l <= +(u, _1), +(u, _1) <= u ---> e_4 = e_12        L: 0; Extension ground terms: 
 [] u = +(u, _1), l <= +(u, _1), +(u, _1) <= u ---> e_4 = e_13           L: 0; Extension ground terms: 
 [] +(u, _1) = +(u, _1), l <= +(u, _1), +(u, _1) <= u ---> e_4 = e_14    L: 0; Extension ground terms: 
 [] +(u, _2) = +(u, _1), l <= +(u, _1), +(u, _1) <= u ---> e_4 = e_15    L: 0; Extension ground terms: 
 [] l = +(u, _2), l <= +(u, _2), +(u, _2) <= u ---> e_5 = e_11           L: 0; Extension ground terms: 
 [] sk_1 = +(u, _2), l <= +(u, _2), +(u, _2) <= u ---> e_5 = e_12        L: 0; Extension ground terms: 
 [] u = +(u, _2), l <= +(u, _2), +(u, _2) <= u ---> e_5 = e_13           L: 0; Extension ground terms: 
 [] +(u, _1) = +(u, _2), l <= +(u, _2), +(u, _2) <= u ---> e_5 = e_14    L: 0; Extension ground terms: 
 [] +(u, _2) = +(u, _2), l <= +(u, _2), +(u, _2) <= u ---> e_5 = e_15    L: 0; Extension ground terms: 
  ---> l <= sk_1                                                         L: 0; Extension ground terms: 
  ---> sk_1 <= +(u, _1)                                                  L: 0; Extension ground terms: 
 e_7 = e_12 --->                                                         L: 0; Extension ground terms: 
  ---> e_9 = e_14                                                        L: 0; Extension ground terms: 
\end{MySmallVerbatim}

\noindent The introduced definitions are replaced by the corresponding congruence
axioms.

\begin{MySmallVerbatim}
Replacing D by N0: 
 This yields 30 clauses.
 +(u, _1) = +(u, _2) ---> e_14 = e_15                                    L: 0; Extension ground terms: 
 u = +(u, _2) ---> e_13 = e_15                                           L: 0; Extension ground terms: 
 u = +(u, _1) ---> e_13 = e_14                                           L: 0; Extension ground terms: 
 sk_1 = +(u, _2) ---> e_12 = e_15                                        L: 0; Extension ground terms: 
 sk_1 = +(u, _1) ---> e_12 = e_14                                        L: 0; Extension ground terms: 
 sk_1 = u ---> e_12 = e_13                                               L: 0; Extension ground terms: 
 l = +(u, _2) ---> e_11 = e_15                                           L: 0; Extension ground terms: 
 l = +(u, _1) ---> e_11 = e_14                                           L: 0; Extension ground terms: 
 l = u ---> e_11 = e_13                                                  L: 0; Extension ground terms: 
 l = sk_1 ---> e_11 = e_12                                               L: 0; Extension ground terms: 
 +(u, _1) = +(u, _2) ---> e_9 = e_10                                     L: 0; Extension ground terms: 
 u = +(u, _2) ---> e_8 = e_10                                            L: 0; Extension ground terms: 
 u = +(u, _1) ---> e_8 = e_9                                             L: 0; Extension ground terms: 
 sk_1 = +(u, _2) ---> e_7 = e_10                                         L: 0; Extension ground terms: 
 sk_1 = +(u, _1) ---> e_7 = e_9                                          L: 0; Extension ground terms: 
 sk_1 = u ---> e_7 = e_8                                                 L: 0; Extension ground terms: 
 l = +(u, _2) ---> e_6 = e_10                                            L: 0; Extension ground terms: 
 l = +(u, _1) ---> e_6 = e_9                                             L: 0; Extension ground terms: 
 l = u ---> e_6 = e_8                                                    L: 0; Extension ground terms: 
 l = sk_1 ---> e_6 = e_7                                                 L: 0; Extension ground terms: 
 +(u, _1) = +(u, _2) ---> e_4 = e_5                                      L: 0; Extension ground terms: 
 u = +(u, _2) ---> e_3 = e_5                                             L: 0; Extension ground terms: 
 u = +(u, _1) ---> e_3 = e_4                                             L: 0; Extension ground terms: 
 sk_1 = +(u, _2) ---> e_2 = e_5                                          L: 0; Extension ground terms: 
 sk_1 = +(u, _1) ---> e_2 = e_4                                          L: 0; Extension ground terms: 
 sk_1 = u ---> e_2 = e_3                                                 L: 0; Extension ground terms: 
 l = +(u, _2) ---> e_1 = e_5                                             L: 0; Extension ground terms: 
 l = +(u, _1) ---> e_1 = e_4                                             L: 0; Extension ground terms: 
 l = u ---> e_1 = e_3                                                    L: 0; Extension ground terms: 
 l = sk_1 ---> e_1 = e_2                                                 L: 0; Extension ground terms: 
\end{MySmallVerbatim}

\noindent Finally we hand over to a prover (Yices is default here).
The program checked earlier that the problem was in a decidable fragment of the
theory of arrays (APF), which is $\Psi$-local. 
Hence Yices' answer can always be trusted irrespective of whether 
the answer is 'satisfiable' or 'unsatisfiable'.


\begin{MyFVerbatim}[frame=lines]
The problem is in APF
Handing over to Yices: 
Total number of clauses: 109.
    unsat
unsat
H-PILoT spent                        0.161975s on the problem.
Of which clausification took         0.006998s.
The prover needed                    0.021996s for the problem.
Total running time:                  0.183971s.
\end{MyFVerbatim}

\subsection{Model generation and visualization}
\label{sec-model-generation}

We illustrate the method for model generation we use on two
examples. 

\subsection*{Example 1: Theory of pointers}

Consider the following problem, specified in H-PILoT input syntax: 

\medskip

\begin{MyFVerbatim}
Base_functions:={(+,2), (-, 2)}
Extension_functions:={(next, 1, 1, pointer), (prev, 1, 1, pointer), 
                      (priority, 1, 1, pointer, real)}
Relations:={(>=, 2)}
Constants:={(null, pointer), (a, pointer), (b, pointer)}

% K
Clauses := (FORALL p). prev(next(p)) = p;
           (FORALL p). --> next(prev(p)) = p;
           (FORALL p). priority(p) >= priority(next(p));

Query := priority(a) = _5; priority(b) = _6; a = prev(b);
        %NOT(a = null); % pivotal for satisfiability
         NOT(b = null);
\end{MyFVerbatim}

\noindent 
We used CVC3 to generate a model for the set of clauses obtained after the 
hierarchical reduction. A partial model is given below (after preprocessing/deleting repetitions):

\begin{MyVerbatim}[frame=lines]
null = a
prev(b) = a
prev(a) = d_5
next(prev(prev(b))) = e_5
next(prev(b)) = d_1
next(b) = a
prev(prev(b)) = d_5
prev(next(prev(b))) = d_5
prev(next(b)) = d_5
prev(next(a)) = d_5
next(a) = d_1
priority(next(a)) = 0
priority(next(prev(b))) = 0
priority(prev(b)) = 5
priority(b) = 6
priority(a) = 5
\end{MyVerbatim}

\medskip
\noindent 
We can make this model total by defining ${\sf next}(x) := {\sf null}$
and ${\sf prev}(y) := {\sf null}$ whenever ${\sf next}(x)$ resp.\
${\sf prev}(y)$ are undefined (cf. Section~\ref{sec-pointers}). The obtained model
can be visualized using Mathematica (this last step is currently
performed separately; it is not yet integrated into H-PILoT, but an
integration is planned for the near future.). The result is presented
below.

\medskip

\noindent \includegraphics[width=12.5cm, height=3.5cm]{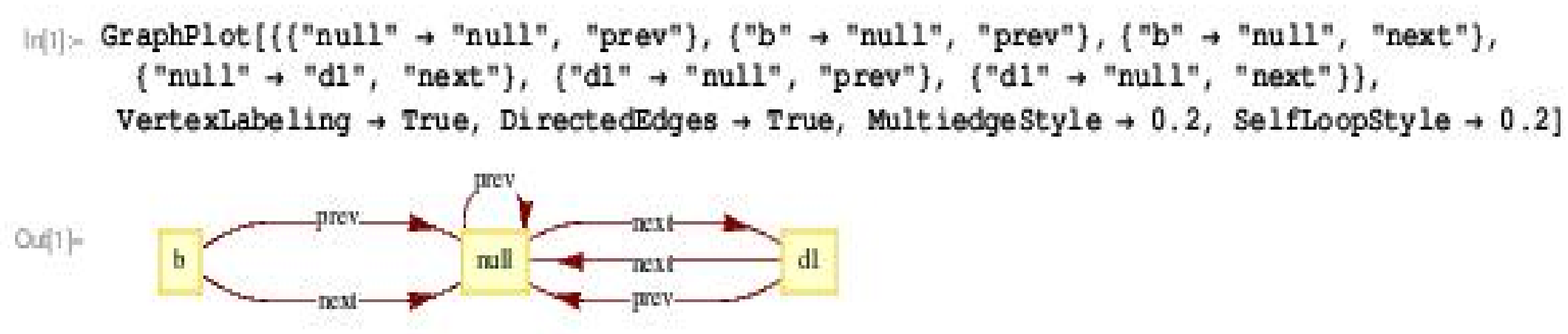}

\subsection*{Example 2: Theory of functions over the real numbers} 

Consider now the following example: decide 
whether
$$ {\sf Mon}_f \cup {\sf Mon}_g \models_{\mathbb R} \forall x, y, z, u, v (
x {\leq} y \wedge z {\leq} y \wedge f(y) {\leq} g(u) \wedge u {\leq} v \wedge u {\leq}  w \rightarrow f(x) {\leq} g(v)).$$
We formulate a satisfiable version, by replacing the argument $x$ in the 
conclusion with a new variable $x_0$.  
The problem obtained this way 
can be formulated as follows in the input format of H-PILoT.

\begin{MyFVerbatim}
Base_functions:={}
Extension_functions:={(f, 1), (g, 1)}
Relations:={(<=, 2)}

Clauses := (FORALL x,y). x <= y --> f(x) <= f(y);
           (FORALL x,y). x <= y --> g(x) <= g(y);

Query := c1 <= d1; c2 <= d1; d2 <= c3; d2 <= c4;
         f(d1) <= g(d2); NOT(f(c0) <= g(c4));
\end{MyFVerbatim}

\noindent CVC3 can be used to generate the following model of the proof task:

\begin{MyVerbatim}[frame=lines]
model:
c0 = 1
c1 = 0
d1 = 0
c2 = 0
c3 = 0
c4 = 1
d2 = 0
g(d2) = 0
f(d1) = 0
g(c4) = 1
f(c0) = 2
\end{MyVerbatim}

\noindent From this partial model, a total model can be constructed 
as explained in \cite{sofronie-ki08}. This model can then be visualized as follows in Mathematica: 

\noindent \includegraphics[width=8.5cm]{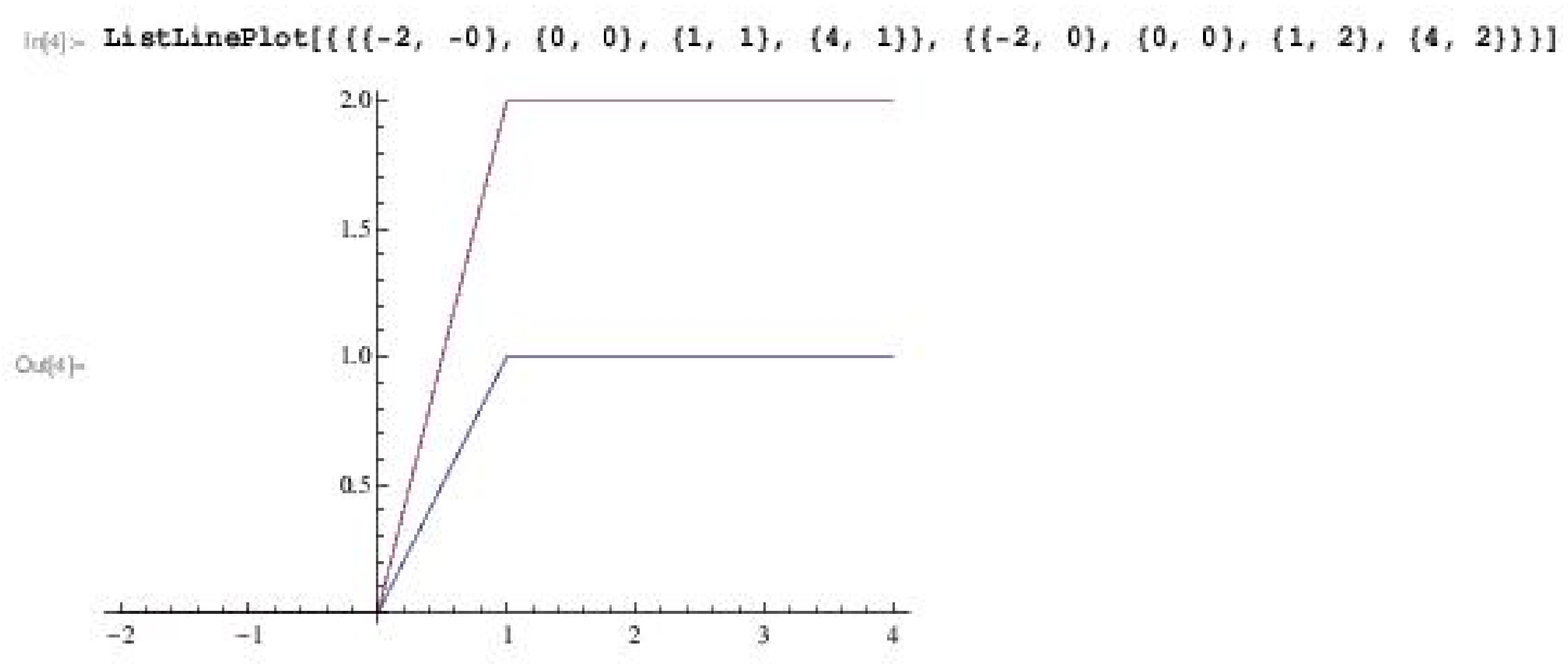}

\vspace{-0.5cm}
\noindent 
Note that if we require differentiability of $f$ and $g$ then 
--  with the completion described in the previous 
example  -- monotonicity of the extensions may not be preserved: 

\vspace{-0.1cm}
\noindent \includegraphics[width=8.5cm]{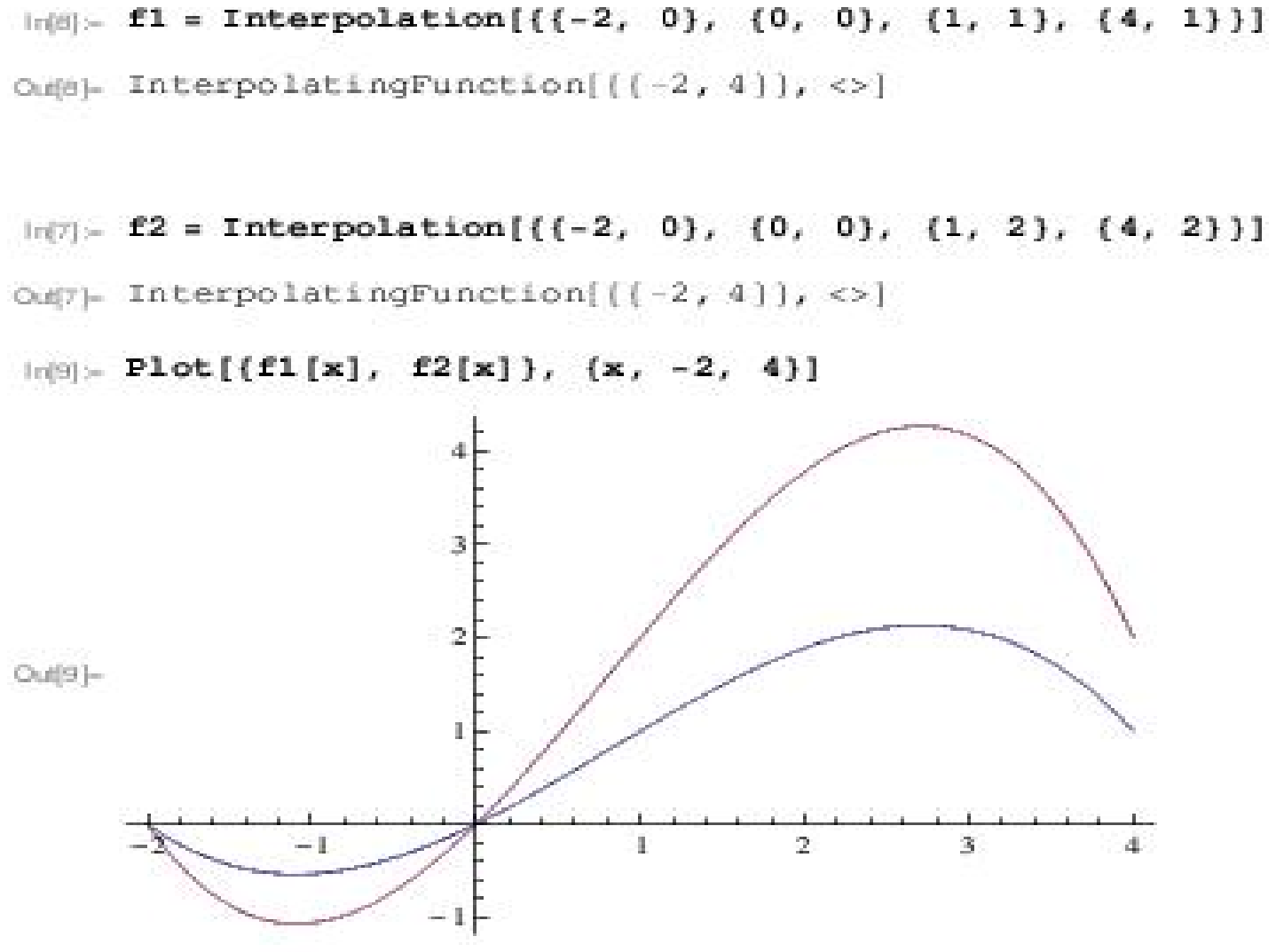}

\vspace{-0.5cm}
\noindent In the example above we can enforce the functions to be linear.
A general study of the properties which can be guaranteed when building 
the models is work in progress.  

\vspace{-0.1cm}
\noindent \includegraphics[width=8.5cm]{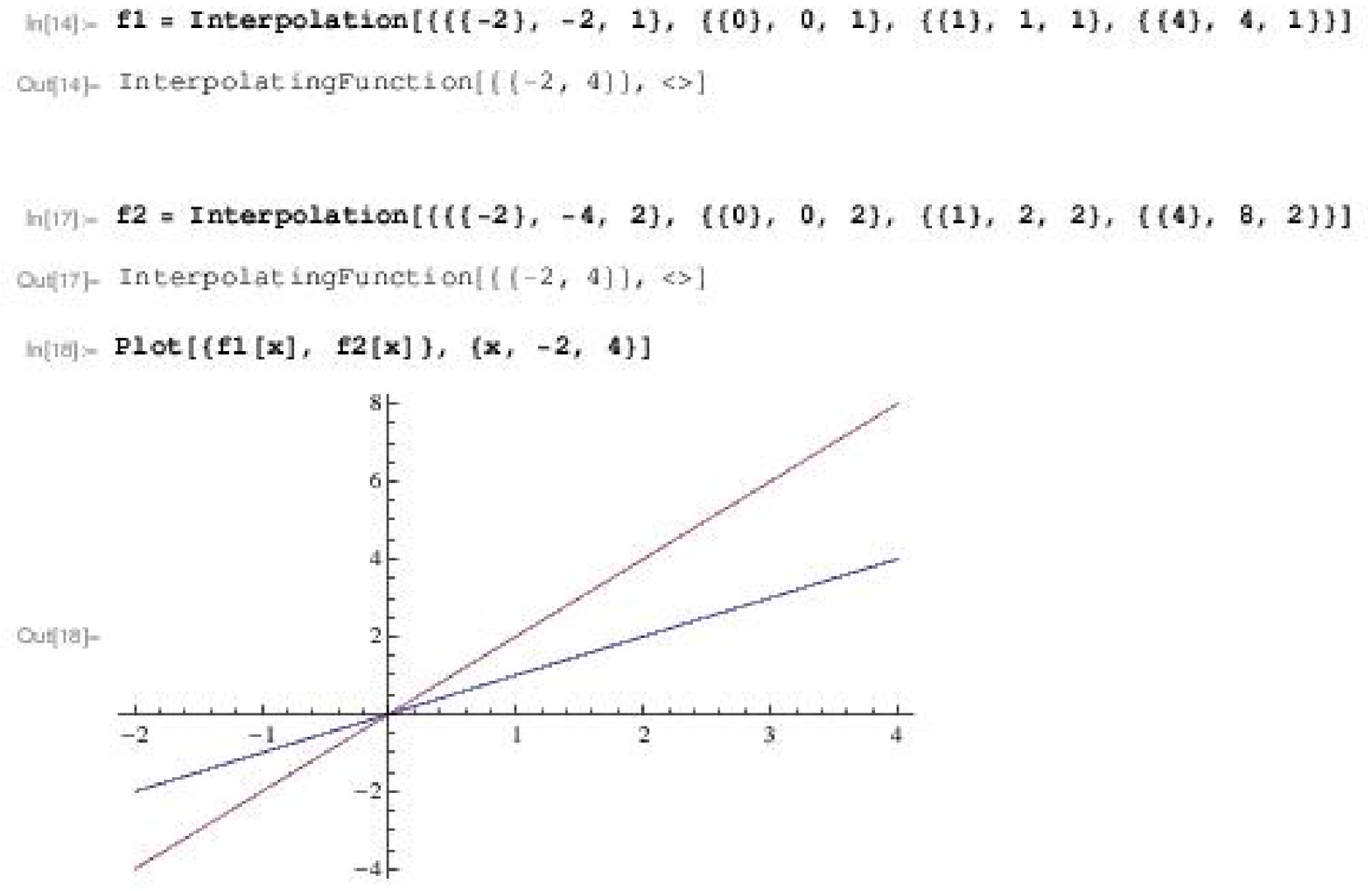}

\subsubsection*{Acknowledgments.} This work was partly
  supported by the German Research
  Council (DFG) as part of the Transregional
  Collaborative Research Center ``Automatic
  Verification and Analysis of Complex
  Systems'' (SFB/TR 14 AVACS, \texttt{www.avacs.org}).


\end{document}